\documentclass[aps,prb,twocolumn,superscriptaddress,showpacs,longbibliography]{revtex4-1}
\usepackage[colorlinks=true,citecolor=blue,linkcolor=blue,breaklinks=true]{hyperref}
\usepackage{amssymb}
\usepackage{graphicx}
\usepackage{amsmath}
\usepackage{mathtools}
\usepackage[export]{adjustbox}
\usepackage{epsfig}
\usepackage{times}
\usepackage{color}
\usepackage{setspace}
\usepackage{calc}
\usepackage{natbib}
\usepackage{bm}

\begin{document}

\newcommand{\rcol} {\textcolor{red}}
\newcommand {\beq} {\begin{equation}}
\newcommand {\eeq} {\end{equation}}
\newcommand {\bqa} {\begin{eqnarray}}
\newcommand {\eqa} {\end{eqnarray}}

\begin{abstract}
The Born-Markov approximation is widely used to study dynamics of open quantum systems coupled to external baths. Using Keldysh formalism, we show that the dynamics of a system of bosons (fermions) linearly coupled to non-interacting bosonic (fermionic) bath falls outside this paradigm if the bath spectral function has non-analyticities as a function of frequency. In this case, we show that the dissipative and noise kernels governing the dynamics have distinct power law tails. The Green's functions show a short time ``quasi'' Markovian exponential decay before crossing over to a power law tail governed by the non-analyticity of the spectral function. We study a system of bosons (fermions) hopping on a one dimensional lattice, where each site is coupled linearly to an independent bath of non-interacting bosons (fermions). We obtain exact expressions for the Green's functions of this system which show power law decay $\sim |t-t'|^{-3/2}$. We use these to calculate density and current profile, as well as unequal time current-current correlators. While the density and current profiles show interesting quantitative deviations from Markovian results, the current-current correlators show qualitatively distinct long time power law tails $|t-t'|^{-3}$ characteristic of non-Markovian dynamics. We show that the power law decays survive in presence of inter-particle interaction in the system, but the cross-over time scale is shifted to larger values with increasing interaction strength.
\end{abstract}

 \title{Power law tails and non Markovian dynamics in open quantum systems: An exact solution from Keldysh field theory}
 
 \author{Ahana Chakraborty and Rajdeep Sensarma}
  \affiliation{Department of Theoretical Physics, Tata Institute of Fundamental
 Research, Mumbai 400005, India.}
 
 \date{\today}
\maketitle
\section{Introduction}

The dynamics of open quantum systems (OQS)~\cite{OQSBook} is the key to some fundamental questions in statistical physics, including issues of emergence of irreversibility and generation of entropy~\cite{irrev_qdyn,*irrev_expt}, generation of quantum entanglement~\cite{Ent_gen} and approach to thermal equilibrium~\cite{ThermalApproach}. It is also crucial to understand the dynamics of a quantum system coupled to an external bath to design and control possible platforms for creating an architecture of quantum computing, like superconducting qubits~\citep{ScQubit1,*ScQubit2}, spin-qubits~\cite{SpQubit}, cavity QED~\cite{CavityQED}, cavity optomechanics~\citep{OptoMech}, quantum dot arrays coupled to cavities ~\cite{takis}, nanowire junctions~\citep{nanowire}, ultracold atomic systems~\cite{coldatom} etc. The phenomenal experimental advance over the past decade in implementing and controlling these platforms has reignited the interest in dynamics of OQS. Additionally, theoretical ideas of bath/dissipation engineering~\citep{bath_engg1} to guide open quantum systems to novel steady states~\cite{steadystate} and using these states as resources in quantum computing~\cite{qcomp_st} also require a deep understanding of behaviour of OQS.

A widely used paradigm to analyze the dynamics of OQS is the Born-Markov approximation, which assumes that (i) coupling of the quantum system to the bath does not change the dynamics of the bath and (ii) the effective reduced dynamics of the system is local in time. This is often presented in the form of a time local (Markovian) quantum master equation (QME)~\cite{qme}, where the positive rates of transition from one configuration of the system to another depends only on the state of the system at that time. Further approximations lead to well known formulations like the Redfield~\cite{Redfield} and Lindblad~\cite{Lindblad1,*Lindblad2} master equations. The other related approach is that of stochastic Schrodinger equations~\citep{qnt_osc_stochastic,stochastic2,*stochastic3}, which are generalizations of familiar Langevin equations in classical systems. 
In systems with short range memory kernels, the Markovian approximation emerges as a coarse grained description of OQS dynamics over a scale $\tau_{course}$ in the limit $\tau_s \gg \tau_{course} \gg \tau_b$. Here $\tau_b$ and $\tau_s$ are the autocorrelation times in the bath and the system. 

Non-Markovian dynamics of OQS has been gaining prominence in recent years~\cite{nonMarkov_Review1,*nonMarkov_Review2}. A broad class of systems, like BEC in trapped ultracold atoms~\cite{nmarkov_bec1,*nmarkov_bec2}, quantum dots coupled to superfluid reservoirs ~\cite{nonmarkov_qdot1,*nonmarkov_qdot2}, nanomechanical oscillators coupled to BEC~\cite{nonmarkov_nmosc1,*nonmarkov_nmosc2}, atoms/impurities coupled to radiation field in photonic crystals~\cite{nmarkov_phot_xtal1,*nmarkov_phot_xtal2} have displayed non-Markovian dynamics in various forms. Non-Markovian dynamics has been key to recent proposals for bath engineering~\cite{bath_engg2}, quantum metrology~\citep{quant_metrology}, and can be used as resources for quantum communication~\cite{nm_qcom} and quantum memory~\cite{nm_mem}. Recent experiments have been successful in tuning the dynamics of an open quantum system from Markovian to non-Markovian by controlling the bath degrees of freedom~\citep{nmarkov_tune_expt}. Non-Markovian dynamics~\citep{nmarkov_stochastic_schrodinger,*nonmarkov_onesite,*twosite_nonmarkov,*nonmarkov_master_1site,*nmarkov_barry,*nonmarkov_qbit,
*segal_nmarkov_powerlaw,*vacchini_nmarkov_powerlaw,*weiss,*DmitryPhysRevB2016,nori_scirep} has been traditionally treated using Nakajima-Zwanzig type master equation \citep{nori,nmarkov_nakazima} as well as effective time convolution-less master equation ~\citep{nmarkov_tlocal,*nmarkov_timelocalmaster,*nmarkov_tlocal_qbit,*nmarkov_tlocal_2level}. In fact, different types of dynamical behaviour or properties of equation of motion is clubbed under the rubric of non-Markovian dynamics with debates over essential definition of \textquotedblleft non-Markovian\textquotedblright ness~\citep{nmarkov_newdef}.

 A canonical model of OQS~\cite{vernon}, consisting of a few bosons (fermions) coupled linearly to non-interacting bath of bosons (fermions), 
has been studied previously using QME~\citep{atomtronics,*rivas,*santos,*luca_feraldi,Abhishek2016} as well as non-equilibrium Greens' function technique~\citep{negf_review,*negf_caroli,*negf_meir,*keldysh_meir_time,*negf_abhishek_diptiman,*negf_photoncavity,*keldysh_scbath_flq,
*keldysh_qdot_scbath,ness_abhishek}. In this paper, we use Schwinger-Keldysh field theory formalism ~\citep{kamenevbook} to study a system bosons(fermions) hoping on a 1d lattice where each site is connected to a non-interacting bosonic (fermionic) bath kept at fixed temperature and chemical potential. Integrating out the bath degrees of freedom, we obtain a description of the effective steady state dynamics of the OQS. We relate the self-energies induced by the bath to the dissipative and noise kernels in a stochastic Schrodinger equation through a saddle point approximation. We show that any non-analyticity in the bath spectral function leads to temporally long range dissipative and noise kernels with power law tails. The exponent of the power law is determined solely by the nature of the non-analyticity and is independent of other microscopic details. Such non-analytic behaviour can arise from a variety of sources like band edges, Van-Hove singularities, Kohn anomalies~\citep{kohn} in phonon spectrum, phase transitions in the bath~\cite{sachdev2007quantum} and non-Fermi liquid~\citep{nonfermiliquid1,nonfermiliquid2} nature of the bath. The power law tail in the kernels preclude the possibility of coarse graining to obtain a Markovian description for the dynamics of the system. We emphasize that, for bosonic baths kept at fixed chemical potentials, i.e when the bath, decoupled from the system has a conserved number/charge, the bath spectrum must be bounded from below and hence its spectral function must be non-analytic at the bottom of the band. Hence non-Markovian dynamics will be ubiquitous in such bosonic systems.
\begin{figure}[t]
\includegraphics[width=0.48\textwidth]{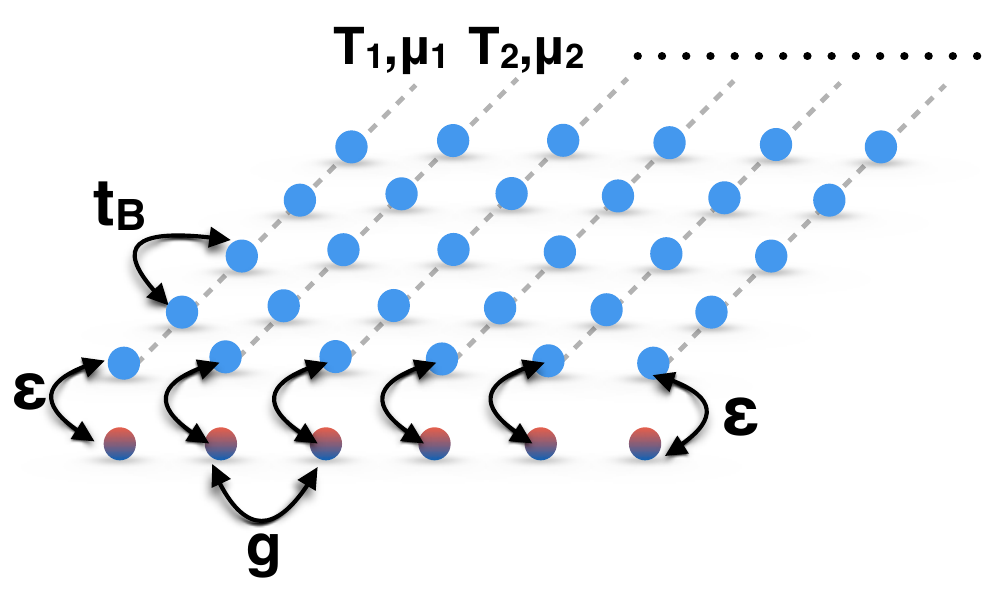}
\caption{Schematic diagram of the open quantum system setup considered in this paper. The red-blue circles denote the system sites, whereas the line of light blue circles denote the semi-infinite bath. Note that the baths are not interconnected to each other and have their independent temperature $T_l$ and chemical potential $\mu_l$. The system sites couple to the first site of the respective bath with scale $\epsilon$. Here $g$ is the hopping in the system, while $t_B$ is the hopping in the bath.}
\label{figure:setup}
\end{figure}

The Green's functions inherit these power law tails along with a short time exponential decay. The crossover from the exponential decay to the power law tail occurs at a time $\tau_0 \sim 1/\epsilon^2$, where $\epsilon$ is the scale of system bath coupling. Thus at very weak system bath coupling the system dynamics appears Markovian for a very long time, and we call this regime ``quasi-Markovian''. With increase in system bath coupling, the power law tail dominates the dynamics and the non-Markovian behaviour is easier to detect in experiments. We have studied both the density and current profile in the system, as well as the unequal time observables like current-current correlators. While the equal time correlators show important quantitative differences from a Markovian calculation, the unequal time correlators inherit the power law tails and can be used as a direct probe of these singularities. 

The features described above are robust, i.e they only require a non-analytic bath spectral function and a linear system-bath coupling. To illustrate these features, we consider a 1d lattice system of bosonic (fermionic) particles with a nearest neighbour hopping, where each site is connected to an independent bosonic (fermionic) bath with its own temperature and chemical potential. We consider a semi-infinite 1-dimensional bath whose spectral function has square root derivative singularity at the band edges. After integrating out the bath, we solve this problem analytically and obtain closed form expressions for Greens' function and the observables mentioned above. We show that the noise and dissipative kernel show a $|t-t'|^{-\frac{3}{2}}$ decay at long times. The Greens' function and current current correlators have a short time exponential decay followed by a power law tail, where the Greens' functions $\sim |t-t'|^{-\frac{3}{2}}$ and current-current correlators $\sim |t-t'|^{-3}$. Since the Green’s functions are known analytically in the model considered here, we can provide insights into the relative strength of exponential and power law decay, as well as the crossover regime between them. We then generalize these relations to different types of non-analyticites and show how the power law changes with the nature of non-analyticity. We thus (i) work out in detailed analytic form, the power law tails and full non-Markovian behaviour of the many body system for a very important and widely used model of OQS and (ii) generalize the description to various other physically relevant non-analyticites in the problem. We also consider the traditional way of studying non-Markovian dynamics in terms of non-exponential (power law) decay of survival probabilities ~\citep{khalfin,knight_refA,oqs_adelfo,parrott_unstable,*seke,*seke1} and relate the power law tails in our formulation to the power law decay of induced populations of modes. Our formalism, where we predict a relation between the power law tails and bath non-analyticites, can thus be used to analyze experiments like Ref[\onlinecite{organic_Monkman}]. While it is known that the non-analytic bath spectral function leads to non-exponential decays~\citep{nori_scirep,nori}, we here provide closed form solution for a number of widely used models of bath spectral function and hence make detailed connection between the nature of the non-analyticity to the exponent of the power laws measured in the observable quantities of the system. 

The equal time correlators like density and current show important quantitative deviations from Markovian description. Our exact solution also allows us to study the evolution of multi-time many body correlators of the open quantum system under the non-Markovian dynamics. These unequal time correlators provide the cleanest observable signal of the non-Markovian dynamics through long time power law tails related to the non-analyticity of the bath spectral function. We have obtained the analytic expressions for the Greens' function of a full 1d chain which provide the starting point for calculating the dynamics of the interacting many-body OQS. We consider the effect of the inter-particle interactions in the system on the dynamics of OQS within mean field theory. We find that, the power law tail survives, while the crossover scale increases with increasing interaction strength. 

We have organized the paper into several sections: In section~\ref{sec:model}, we discuss a model of non-interacting bosons linearly coupled to a bosonic bath. We set up the Keldysh formalism and sketch the steps taken to calculate the Green's function and observables in the resulting OQS. In section~\ref{sec:diss_noise}, we connect the self energies induced by integrating out the bath variables with the dissipative and noise kernels in an equation of motion approach and discuss the power law kernels induced by the non-analyticities in the bath spectral function. In section~\ref{sec:green_chain}, we solve the Dyson equation and find the analytic Green's function for the linear chain. We discuss the non-analytic structure of the Green's function in frequency space, and its behaviour in real time. In section ~\ref{sec:observables}, we consider equal time and unequal time observables and show how the unequal time correlators show the signature of the power law tails. We also relate our formalism with the more standard approach of studying non-Markovian dynamics in terms of decay of survival probabilities. In section~\ref{sec:fermions}, we generalize the results to the case of a fermionic system coupled to a fermionic bath, and show that similar results hold in this case. Finally, in section~\ref{sec:interactions}, we consider the effect of inter-particle interactions in the system on the power law tail and the crossover time-scale. 

\section{Keldysh theory for Bosonic system}\label{sec:model}
 We consider a system of 1-dimensional lattice of non-interacting bosons (representing arrays of oscillators) hopping between nearest neighbours with an amplitude g (representing coupling between successive oscillators).  
The fermionic version of this model, which can represent quantum dot arrays, will be taken up in section~\ref{sec:fermions}.
  Each site, $l$, of the lattice is coupled to an independent bosonic bath with  a temperature $T_l$ and  a chemical potential $\mu_l$. The setup is schematically shown in Fig.~\ref{figure:setup}. We will be interested in the steady state response of the system to different profiles of $\mu_{l}$ and $T_{l}$.

We describe each bath by a semi-infinite 1 dimensional lattice of non-interacting bosons with nearest neighbour hopping $t_B$ and  assume each system site is coupled locally to the first site of the corresponding bath, as shown in Fig.~\ref{figure:setup}. The system bath coupling ,  controlled by the scale $\epsilon$ , is linear in both the system and bath degrees of freedom . The total Hamiltonian of the system ($H_{s}$) , the baths ($H_{b}$) and system bath coupling ($H_{sb}$) are then given by,
\begin{eqnarray}\label{model_Ham}
\nonumber H_{s}& =& -g\sum_{l=1}^{N}  a_{l}^{\dagger}a_{l+1} + h.c  ~~ and ~~ H_{sb} = \epsilon\sum_{l=1}^{N} a_{l}^{\dagger} b^{ (l)}_{1} + h.c \\
H_{b} &=& -t_B\sum_{l=1}^{N} \sum_{s=1}^{\infty}  b^{(l)\dagger }_{s} b^{ (l)}_{s+1}  +h.c. 
\end{eqnarray}
where $a_{l}^{\dagger}$ creates a boson at site $l$ of the system, $b_{s}^{(l)\dagger }$ is the bosonic creation operator at site $s$ of $l^{th}$ bath. It is useful to rewrite the bath degrees of freedom in terms of the eigenoperators  $B^{(l)\dagger}_{\alpha}$ which diagonalize the bath Hamiltonian. 
\begin{eqnarray}\label{bath_eigen}
H_{b}  =\sum_{l,\alpha}\Omega_{\alpha} B^{(l)\dagger}_{\alpha} B^{ (l)}_{\alpha} ~,~  H_{sb}  =\epsilon \sum_{l,\alpha} \kappa_{\alpha} B^{(l)\dagger}_{\alpha} a_{l} + h.c 
\end{eqnarray}
where $\Omega_\alpha$ is the energy of the eigenmode $\alpha$ and $\kappa_\alpha$ is its amplitude on the first site of the bath.
\begin{figure}[t]
\centering
\includegraphics[width=0.3\textwidth]{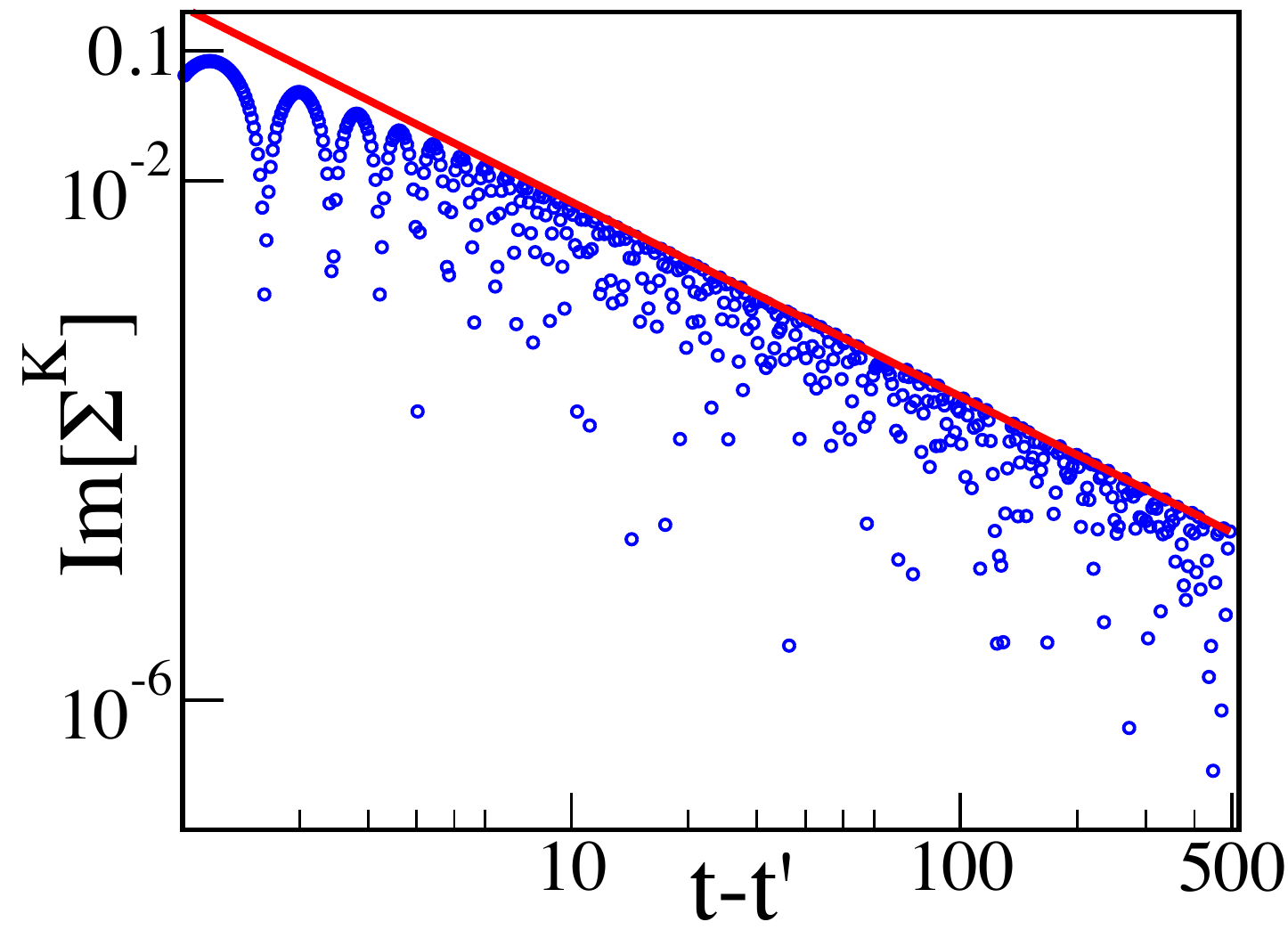} 
\caption{The imaginary part of the noise kernel (Keldysh self energy) $\Sigma_l^K(t-t')$ of a site $l$ coupled to a bath with $T_l=0.625t_B$ and $\mu_l =-2.25 t_B$ as a function of $|t-t'|$ on a log-log plot. The system bath coupling in this case is $\epsilon = 0.3 t_B $. Note that the absolute value of the self energy is plotted in this case. The solid line is the analytic answer for the envelop of the leading power law $\sim |t-t'|^{-3/2}$ decay of the kernel. We use $t_B = 2$ to set units of $t-t'$ and $\Sigma^{K}(t-t')$.}
\label{figure:noise_kernel}
\end{figure}
We use Schwinger Keldysh functional integral formalism ~\cite{kamenevbook} to study the effective non-unitary dynamics of the OQS in the steady state. The field theoretic technique yields exact results for  arbitrary parameter values in non-interacting system, and is not restricted to weak system bath couplings. The Keldysh approach constructs a path integral representation of the dynamics of the density matrix . It requires two copies of fields at each instant of time $t$, namely $\phi_{+}(t)$ and $\phi_{-}(t)$, corresponding to the forward and backward evolution inherent in $\rho(t) = \mathcal{U}(t,-\infty)\rho(-\infty)\mathcal{U}^{\dagger}(t,-\infty)$. Here $U(t)$ is the time evolution operator for the system as well as the baths and $\rho(-\infty)$ is the initial density matrix, which is factorizable into system and bath density matrices. It is customary to work with the symmetric or classical $ \phi_{cl}=(\phi_{+}+ \phi_{-} )/\sqrt{2} $ and anti-symmetric or quantum $ \phi_{q}=(\phi_{+} - \phi_{-})/\sqrt{2}$ fields. Using $\phi^{(l)}$ and $ \psi^{(l)}_{\alpha} $ as the fields corresponding to $a_{l}$ and $B^{(l)}_{\alpha}$ , the Keldysh action for the system $S_{s}$, baths $S_{b}$, and the system bath couplings $S_{sb}$ is given by
\begin{equation}
\nonumber S_{s} = \sum_{l,l'}\int d\omega
\phi^\dagger_l(\omega)
    \left[ {\begin{array}{cc}
   0 & G^{-1A}_{0}(l,l',\omega)\\ 
    G^{-1R}_{0}(l,l',\omega)& G_{0} ^{-1K}(\omega)~\delta_{l,l'} \\
  \end{array} } \right]\phi_{l'}(\omega)
  \label{action_system}
\end{equation}
\begin{equation}
\nonumber S_{b} = \sum_{l,\alpha}\int d\omega~
\psi^\dagger_{l\alpha}(\omega)
    \left[ {\begin{array}{cc}
   0 & \omega-\Omega_{\alpha}-\mathbf{i}\eta \\ 
    \omega-\Omega_{\alpha}+\mathbf{i}\eta& 2\mathbf{i}\eta~F_{l}(\Omega_{\alpha}) \\
  \end{array} } \right]
\psi_{l\alpha}(\omega)
  \label{action_bath}
\end{equation}
\begin{equation}
S_{sb} = -\epsilon\sum_{l,\alpha}\int d\omega~ \kappa_{\alpha}
\psi^\dagger_{l\alpha}(\omega)\hat{\sigma}_1\phi_l(\omega) +h.c
  \label{action_sb}
\end{equation}
where $\sigma^1$ is the Pauli matrix encoding the Keldysh rotation and $\phi^\dagger_l=[\phi^{*(l)}_{cl}~,~ \phi^{*(l)}_{q}] $, $\psi^\dagger_{l\alpha}=[\psi^{*(l)}_{cl,\alpha}~,~\psi^{*(l)}_{q,\alpha}]$ and  $G^{-1R/A}_{0}(l,l',\omega)=(\omega\pm\mathbf{i}\eta)~\delta_{l,l'}+g~\delta_{l,l'\pm1}$, and $\eta \rightarrow 0^{+}$ ~\cite{g0k}. Here $F_{l}(\omega)= \coth \Big(\frac{\omega-\mu_{l}}{2T_{l}}\Big)$ is related to the distribution function in the $l^{th}$ bath. We assume that all the baths remain in thermal equilibrium with the bath Greens' functions given by
\begin{eqnarray}
\nonumber G^{R/A}_{b}(\alpha;\omega) &=& \frac{1}{\omega-\Omega_{\alpha} \pm \mathbf{i}\eta}\\
 G^{K}_{b}(\alpha,l;\omega) &=& -2\pi \mathbf{i}~ \delta(\omega- \Omega_{\alpha}) ~\coth \Big(\frac{\omega-\mu_{l}}{2T_{l}}\Big)
\label{greensfn_eqlbm}
\end{eqnarray}
Since the action (\ref{action_bath}) is quadratic in $\psi^{(l)}_{\alpha}$, we can integrate out the bath degrees of freedom to obtain the effective dissipative action for the open quantum system which describes the non-unitary dynamics of the system coupled to the bath, 
\begin{widetext}
\begin{equation}
S_{oqs} = \sum_{l,l'}\int  d\omega
\phi^\dagger_l(\omega)
    \left[ {\begin{array}{cc}
   0 &~~ G^{-1A}_{0}(l,l',\omega)-\Sigma^{A}(w)~\delta_{l,l'}\\ 
    G^{-1R}_{0}(l,l',\omega)-\Sigma^{R}(w)~\delta_{l,l'}& -\Sigma^{K}_{l}(\omega)~\delta_{l,l'} \\
  \end{array} } \right]
\phi_{l'}(\omega),
  \label{action_diss}
\end{equation}
\end{widetext}
where $\hat{\Sigma}_{l}(\omega)=  \sum_{\alpha}|\kappa_{\alpha}|^{2}~ \epsilon^{2}~\hat{\sigma}_{1} \hat{G}_{b}(\alpha,l;\omega) \hat{\sigma}_{1}$. Here, integrating out the bath induces a finite Keldysh component of the self-energy matrix, $\Sigma^{K}_{l}(\omega)$, which is purely imaginary. Thus, although we start from unitary description of the combined system, the effective dynamics of the OQS after tracing over bath degrees of freedom, governed by the action ($\ref{action_diss}$), is non-unitary. The summation over the bath eigenmodes makes the self-energy matrix a function only of the  bath spectral function
 \begin{equation}\label{bathspec_def}
 J(\omega) = 2 \pi \sum_{\alpha} |\kappa_{\alpha}|^{2} \delta(\omega-\Omega_{\alpha}) ,
 \end{equation}
and the distribution function $F_{l}(\omega)$ . All the results of this paper will depend on properties of $J(\omega)$ and not on other microscopic details of the baths. In that sense, we can simply think of different kinds of baths being represented by different $J(\omega)$, and the results obtained here are applicable to a large class of systems . For our particular case, we will mainly focus on $1-$d semi-infinite bath given by the Hamiltonian (\ref{bath_eigen}) which is diagonalized in the quasi-momentum basis $\alpha$ with the eigen-energies, $\Omega_{\alpha} = -2t_B \cos(\alpha)$. In this formalism, baths are assumed to have infinite number of degrees of freedom so that their energy spectrum is continuous and  $\kappa_{\alpha}$ is proportional to the eigen-function of $H_b$ at the first site of the bath, $\kappa_{\alpha} = \sqrt{2/\pi} \sin(\alpha) $. Using equation (\ref{bathspec_def}) we then get,
\beq
 J(\omega) = \Theta(4t_B^2-\omega^2)\frac{2}{t_{B}} \sqrt{1-\frac{\omega^2}{4t^2_{B}}},
\label{Jomega_ourbath}
\eeq
 has a square root derivative singularity at the band edges $\omega =\pm 2t_B$ where $|2 t_B| < |\mu_l | ~\forall ~ l ~,$ to avoid BEC forming in the baths. We discuss the cases of some other physically relevant $J(\omega)$ in section \ref{sec:diss_noise}.  Writing the components of self energy matrix explicitly, we have
\begin{eqnarray}
\label{selfenergy:freq}
\Sigma^{R}(\omega)  &=&   -\epsilon^{2} \int \frac{d\omega'}{2\pi}~\frac{J(\omega')}{\omega'-\omega-i\eta} \nonumber\\
\Sigma^{K}_{l}(\omega) &=&  -\mathbf{i}\epsilon^2 J(\omega)~ \coth\left[\frac{\omega-\mu_{l}}{2T_{l}}\right]    
\end{eqnarray}
Inverting the kernel in the action (\ref{action_diss}) we obtain exact expressions for the 1-particle Greens' function of the bosons. Using this, we calculate equal time observables (e.g. current, occupation number ) as well as unequal time observables (e.g. current current correlation functions) in the steady state. Green's functions and current-current correlators in the steady state show power law tails indicative of non-Markovian dynamics of the system. In the next section, we work in real time domain and connect the Keldysh formalism with the widely used stochastic Schrodinger equation~\citep{qnt_osc_stochastic} formalism.

\begin{table*}[t!]
  \centering
   \begin{tabular}{|c||c||c||c|}
          \hline
 Physical model &Form of $J(\omega)$ &$\Sigma^{R}(t-t')  ~\sim$ & Asymptotic limit  \\
          \hline
& & & \\
Semi-infinite 1-d bath &  $\Theta(4t_B^2-\omega^2)\frac{2}{t_{B}} \sqrt{1-\frac{\omega^2}{4t^2_{B}}}$ & $\frac{\mathcal{J}_{1}(2t_B |t-t'|)}{|t-t'|}$ & $|t-t'|^{-\frac{3}{2}}$
       \\
& & & \\
          \hline
& & & \\
        Infinite 1-d bath &  $\Theta(4t_B^2-\omega^2)\frac{2}{t_{B}} \frac{1}{\sqrt{1-\frac{\omega^2}{4t^2_{B}}}}$ & $\mathcal{J}_{0}(2t_B |t-t'|)$ & $|t-t'|^{-\frac{1}{2}}$ 
  \\ 
& & & \\
 \hline
& & & \\
Ohmic,Sub and Super Ohmic bath&~ $\Theta(\omega) \frac{\omega^{x} }{\omega_{c}^{x+1}} ~exp(-\frac{\omega}{\omega_{c}}) $ & $\frac{\Gamma[1+x]}{[1-\mathbf{i}~w_{c} (t-t')]^{x+1}} $
      & $|t-t'|^{-x-1}$ 
   \\ 
& & & \\
 \hline
& & & \\
 2-d square lattice bath &  $\frac{1}{w_{c}} \log \left( \frac{|w-w_{0}|}{w_{c}}\right)$ &$-~e^{\mathbf{i} w_{0}(t-t')}~ \frac{ sign (t-t')}{t-t'} $ & $|t-t'|^{-1}$  \\
& & & \\ 
           \hline
   \end{tabular}
   \caption{ Leading power law decay in dissipative kernel ($\Sigma^R(t-t')$) at large $t-t'$ for different non-analytic bath spectral functions $J(\omega)$.}
    \label{supp:table:power_law}
\end{table*}

\section{Power law tail in Dissipative and Noise Kernels}\label{sec:diss_noise}
The Keldysh field theory description obtained in the previous section can be connected to a stochastic Schrodinger equation with non-local memory kernels for dissipation and noise in the system given by the self-energies $\Sigma^{R}$ and $\Sigma^{K}_{l}$. To see this, we write the effective Keldysh action for the OQS in real time,
\begin{equation}
S = \sum_{l,l'}\int d t ~dt' \phi^\dagger_l(t)
    \left[ {\begin{array}{cc}
   0 & G^{-1A}_{ll'}(t,t')\\ 
     G^{-1R}_{ll'}(t,t' )& -\Sigma^{K}_{l}(t,t')\delta_{l,l'} 
  \end{array} } \right]
\phi_{l'}(t')
  \label{action_diss_t}
\end{equation}
where $G _{l,l'}^{-1R}(t,t') =\delta(t-t') \big(\mathbf{i} \partial_{t} \delta_{l,l'} +g \delta_{l,l'\pm1} \big) -\Sigma^{R}(t-t')\delta_{l,l'}$. The terms quadratic in $\phi_q$ are first converted into terms linear in $\phi_{q}$ (source terms) by a Hubbard Stratanovich transformation with an auxiliary field $\zeta_{l}(t)$. The Keldysh partition function is then given by $\mathcal{Z}=\int \mathcal{D}[\zeta^{*}\zeta]~F(\zeta^{*}\zeta)\int\mathcal{D}[\phi^{*}\phi]e^{\mathbf{i} S(\phi^{*},\phi,\zeta^{*},\zeta)}$, where
\begin{eqnarray}
S = \int dt \int dt'\sum_{l,l'} [ \phi^{*(l)}_{q} (t)~ G^{-1R}_{l,l'}(t,t' ) \phi^{(l)}_{cl} (t')+ h.c ]\nonumber \\
+ \int \limits_{-\infty}^{\infty} dt ~\sum_{l} \Big[ \phi^{*(l)}_{q} (t) \zeta_{l}(t) + \zeta^{*}_{l}(t) \phi^{(l)}_{q} (t)\Big]~~~~\text{and}~~~~~~~~~~~~\nonumber\\
F(\zeta^{*}\zeta) = e^{\mathbf{i} \int \limits_{-\infty}^{\infty} \int \limits_{-\infty}^{\infty} dt~dt'~\zeta^{*}_{l}(t)[\Sigma^{K}_{l}(t,t')]^{-1}\zeta_{l}(t')}~~~~~~~~~~~~~~~~~~~~~~~~
\end{eqnarray}
The equation of motion, obtained from classical saddle point condition $\frac{\partial \mathcal{S}}{\partial \phi_{q} }\Big \lvert_{\phi_{q} = 0}  = 0$, is 
\begin{equation}
\label{eom}
\mathbf{i}\partial_{t}  \phi^{(l)}_{cl}(t) - \int dt'~\Sigma^{R}_{l}(t,t') \phi_{cl}^{(l)}(t') +g \phi_{cl}^{(l)\pm1}(t)  = \zeta_{l}(t),
\end{equation}
where $\zeta_{l}(t)$ is a complex random field with a non-local but Gaussian distribution, $\left\langle\zeta^{*}_{l}(t)\zeta_{l'}(t') \right\rangle= -\mathbf{i}\delta_{l,l'} \Sigma^{K}_{l}(t,t')$. The imaginary part of the retarded self energy is thus the dissipative kernel, while the Keldysh self energy is the noise correlation kernel. The self energies in real time are related to the bath spectral function by 
\begin{eqnarray}
\Sigma^R(t-t') &=&-\mathbf{i} \epsilon^2\Theta(t-t') \int \frac{d\omega}{2\pi} J(\omega)e^{-\mathbf{i}\omega(t-t')}\\
\nonumber\Sigma^K_l(t-t') &=&-\mathbf{i} \epsilon^2 \int \frac{d\omega}{2\pi} J(\omega)\coth \left[\frac{\omega-\mu_l}{2T_l}\right] e^{-\mathbf{i}\omega(t-t')}\label{sig:t}
\end{eqnarray}
We note that for any $J(\omega)$ which is even in $\omega$, the retarded self energy is purely imaginary and hence dissipative in nature. For the bath spectral function that we consider ( eqn.~\ref{Jomega_ourbath}), 
\beq
\Sigma^{R}(t,t')= -\mathbf{i}~\Theta(t-t')\frac{\epsilon^{2}}{t_{B}}\frac{\mathcal{J}_{1}(2t_{B}|t-t'|)}{|t-t'|},
\label{sigR_exact}
\eeq
where $\mathcal{J}_{1}(x)$ is the Bessel function of first order. It is clear from the above exact expression, that $\Sigma^{R}(t-t') \sim -\mathbf{i} \epsilon^2[t_{B} (t-t')]^{-\frac{3}{2}}\cos [2 t_{B}(t-t')-3\pi/4]$ for $|t-t'|\rightarrow \infty$ i.e the dissipative kernel decays slowly as a power law in the long time limit. Hence, there is no time scale in the system over which one can coarse grain to obtain an effective local description. The dynamics is thus essentially non-Markovian. Although a closed form expression for $\Sigma^{K}(t-t')$ cannot be obtained, it can be shown (see Appendix A) that at long times the envelop of $\Sigma^{K}
\sim - \epsilon^{2}(2/\pi)^{1/2}~|2t_B (t-t')|^{-3/2} \left ( \coth\left[(2t_B-\mu)/2T\right]   \mp   \coth\left[(2t_B+\mu)/2T  \right] \right)$ where the $-(+)$ sign corresponds to its real (imaginary) part. Hence it has the same $\sim |t-t'|^{-3/2}$ tail as $\Sigma^{R}$. In Fig.~\ref{figure:noise_kernel}, we plot the imaginary part of the Keldysh self energy of a site coupled to a bath with temperature $T/t_B=0.625$ and chemical potential $\mu/t_B=-2.25$ as a function of $|t-t'|$ on a log-log plot. The analytic envelop of $\Sigma^{K}$ is plotted in the same figure with a solid line. It matches with the numerically obtained results remarkably well. Here, the chemical potential is kept below the band bottom to avoid the pathological case of non-interacting BEC in the bath. We note that in the limit $\mu \rightarrow -2t_B$, i.e a BEC transition is approached, the $\Sigma^{K}(\omega)$ has a square root divergence at $\omega = \pm 2t_B$ and the power law tail will change to $|t-t'|^{-1/2}$.

The origin of the power law tail can be traced back to the fact that the bath spectral function $J(\omega)$ is non-analytic at $\omega = \pm 2 t_{B}$. This is similar to Friedel oscillations~\citep{mahan2013many} or RKKY interaction for impurities in a Fermi gas, where non-analyticity of the polarization function leads to power law decays in space. We will focus on a particular model of OQS with square root derivative singularity at the band edge of the density of states of the bath in most of the parts of this paper, which will allow us to obtain analytic expressions for various quantities. However, the relation between power law tails and singularities is very robust, and we will work out the power law exponents for a different types of non-analytic spectral functions. For example, one can consider the arrangement where each lattice site is coupled to 1-dimensional bath, where the coupling is to a central site of the bath (infinite 1d bath) as opposed to the one end of the bath (semi-infinite 1d bath). In this case, the bath spectral function is, $J(\omega) = \Theta(4t_B^2-\omega^2)(2/ t_{B}) \left[ 1-\omega^2/(4t^2_{B})\right]^{-1/2} $. Using the same formalism, the analytical expression for $\Sigma^{R}(t,t')= -\mathbf{i}~\Theta(t-t')~2\epsilon^{2}\mathcal{J}_{0}(2t_{B}|t-t'|)$, which scales as $|t-t'|^{-1/2}$ for large $|t-t'|$. Another commonly used spectral function~\citep{nori} is $J(\omega)=\Theta(\omega) \frac{\omega ^{x}}{\omega_{c}^{x+1}} ~exp(-\frac{\omega}{\omega_{c}})$ which is a prototype of ohmic $(x=1)$, sub-ohmic $(x<1)$, super-ohmic $(x>1)$ baths respectively. In this case, we obtain closed form analytic solutions $\Sigma^{R}(t,t')=-\mathbf{i}~\Theta(t-t')~\epsilon^{2}\Gamma[1+x]/(2\pi) \left[1+\mathbf{i}~w_{c} (t-t') \right]^{-(x+1)} \sim |t-t'|^{-x-1} $ in the long time limit. Finally we consider the spectral density of a 2d square lattice which has a logarithimic Van-Hove singularity (not coming from the band edge) of the form $J(\omega) \sim 1/w_{c} \log \left( |w-w_0|/w_{c}\right)$ as $\omega \rightarrow \omega_0$. In this case, we obtain $\Sigma^{R}(t,t') \sim e^{\mathbf{i} w_{0}(t-t')}~ sign (t-t')/(t-t') $ which decays as power law $\sim |t-t'|^{-1}  $. This provides a clear exposition of how the power law exponent is related to the nature of non-analyticity in the bath spectral function. We summarize these results in Table (\ref{supp:table:power_law})  where the exact and asymptotic forms are tabulated. 

For bosonic baths, which can have a chemical potential, $H_B$ must have a conserved particle number. In that case, the spectrum of $H_B$ must be bounded from below to avoid pathologies of condensation at infinitely negative energies and at least one non-analytic point is guaranteed for the bath spectral function. Non- Markovian dynamics is thus ubiquitous for systems linearly coupled to such bosonic baths. Non analytic $J(\omega)$ can occur due to a variety of reasons: e.g. Van-Hove singularities (both bosons and fermions), Kohn anomalies (phonons)~\citep{kohn}, non Fermi liquids (fermions) ~\citep{nonfermiliquid1,*nonfermiliquid2}, critical fluctuations near a quantum phase transitions (bosonic modes) ~\citep{sachdev2007quantum} etc. The non-Markovian dynamics can then be used to detect the presence and nature of these non-analyticities. If the bath spectral function has sharp but non-singular features, the kernel will be an approximate power law for a large intermediate time period, before showing exponential decay on the time scale at which the singularity is smoothed out. 

In the next section, we obtain analytic expression for the Green's function of a linear chain, which we will later use to calculate correlation functions in the OQS.
\begin{figure}[t]
 \centering
 \includegraphics[width=0.238\textwidth]{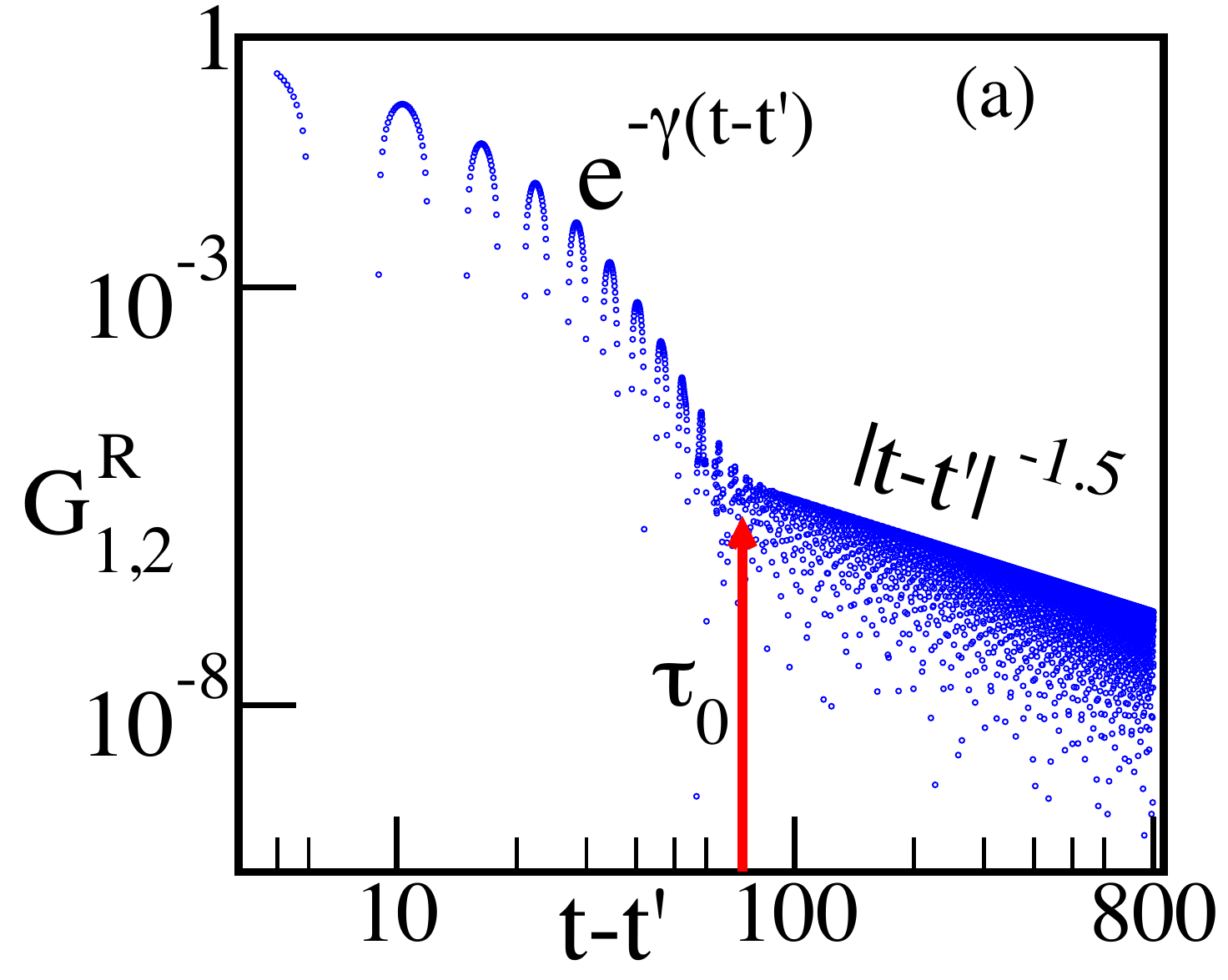}
 \includegraphics[width=0.238\textwidth]{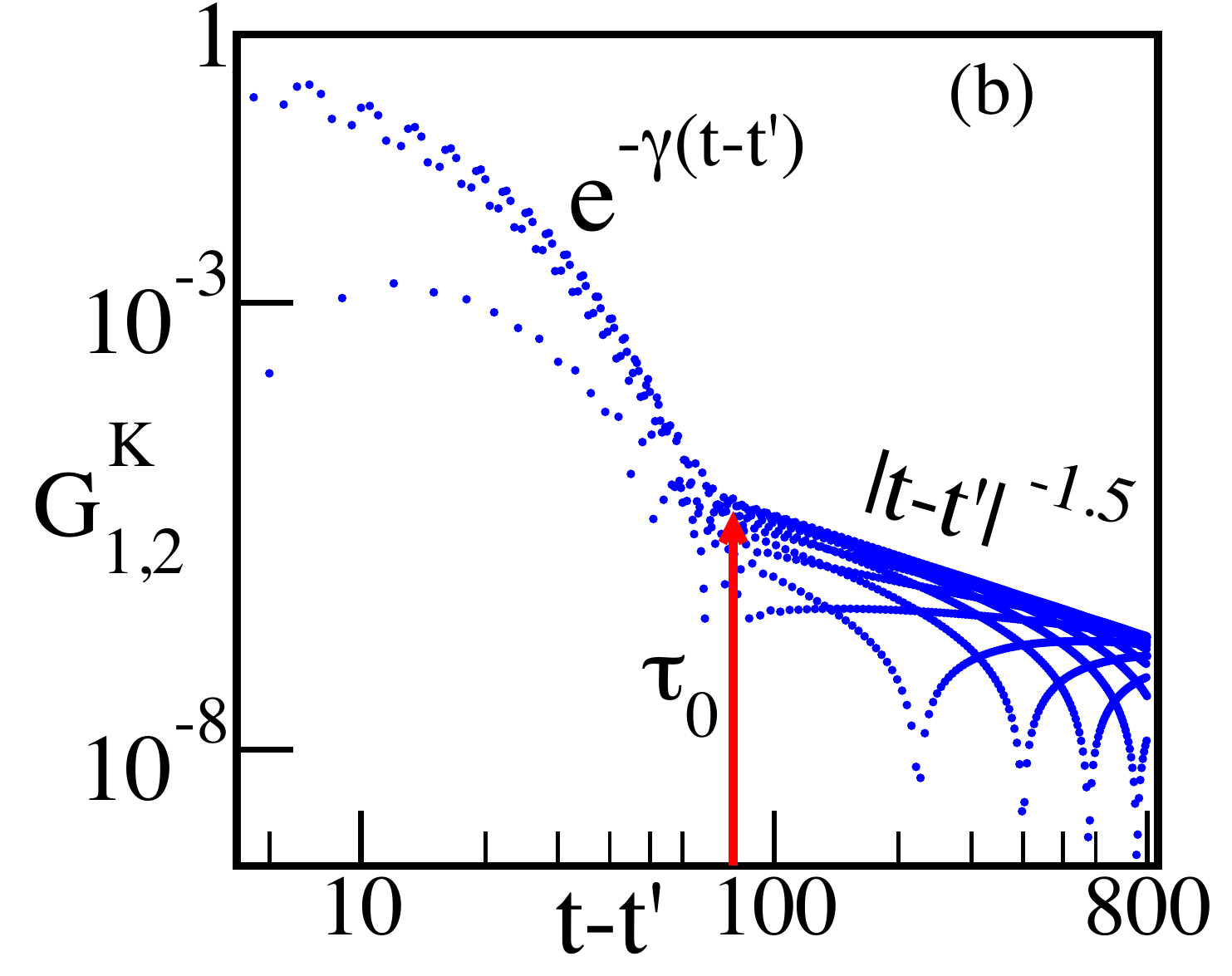}
 \includegraphics[width=0.238\textwidth]{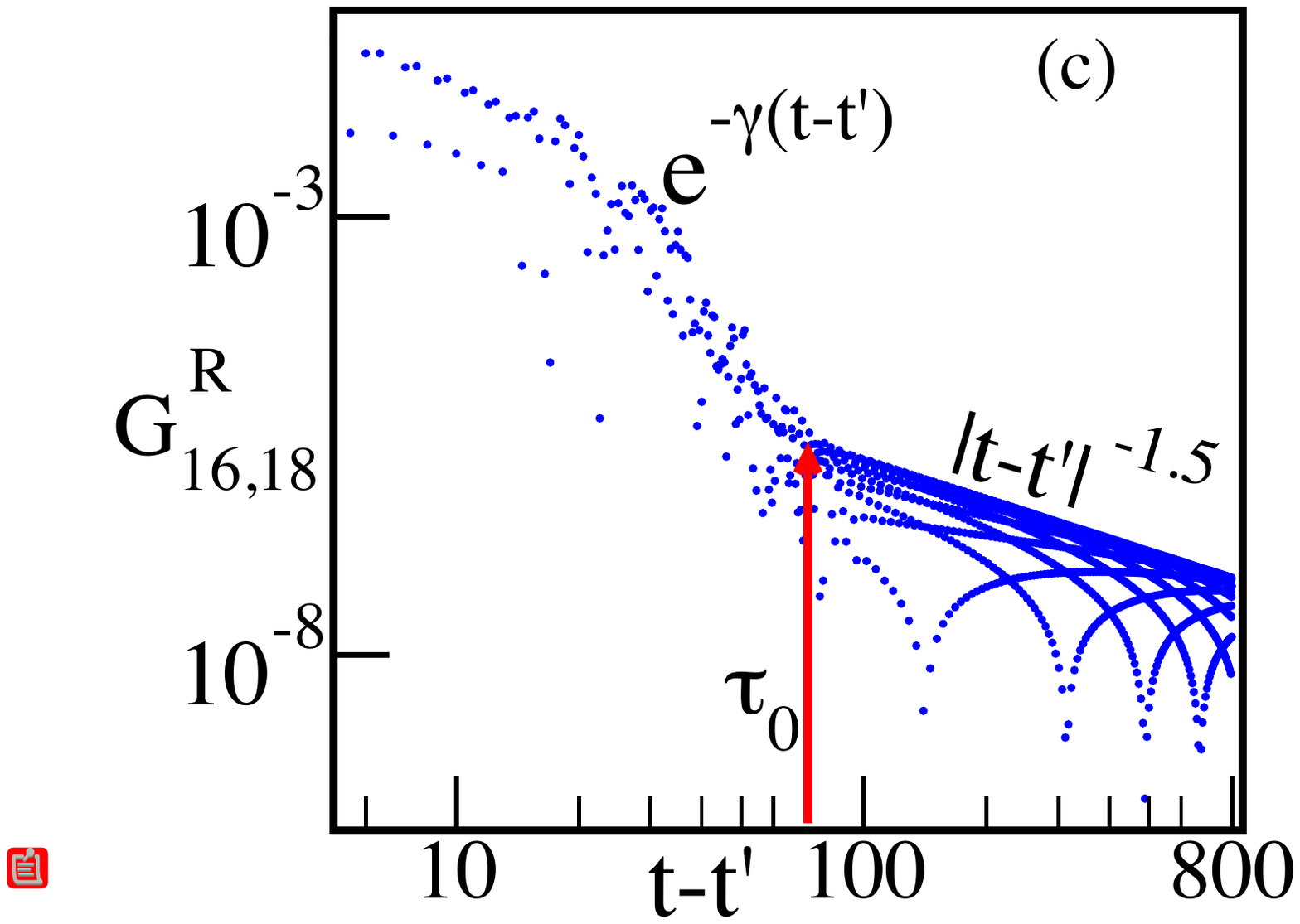}
 \includegraphics[width=0.238\textwidth]{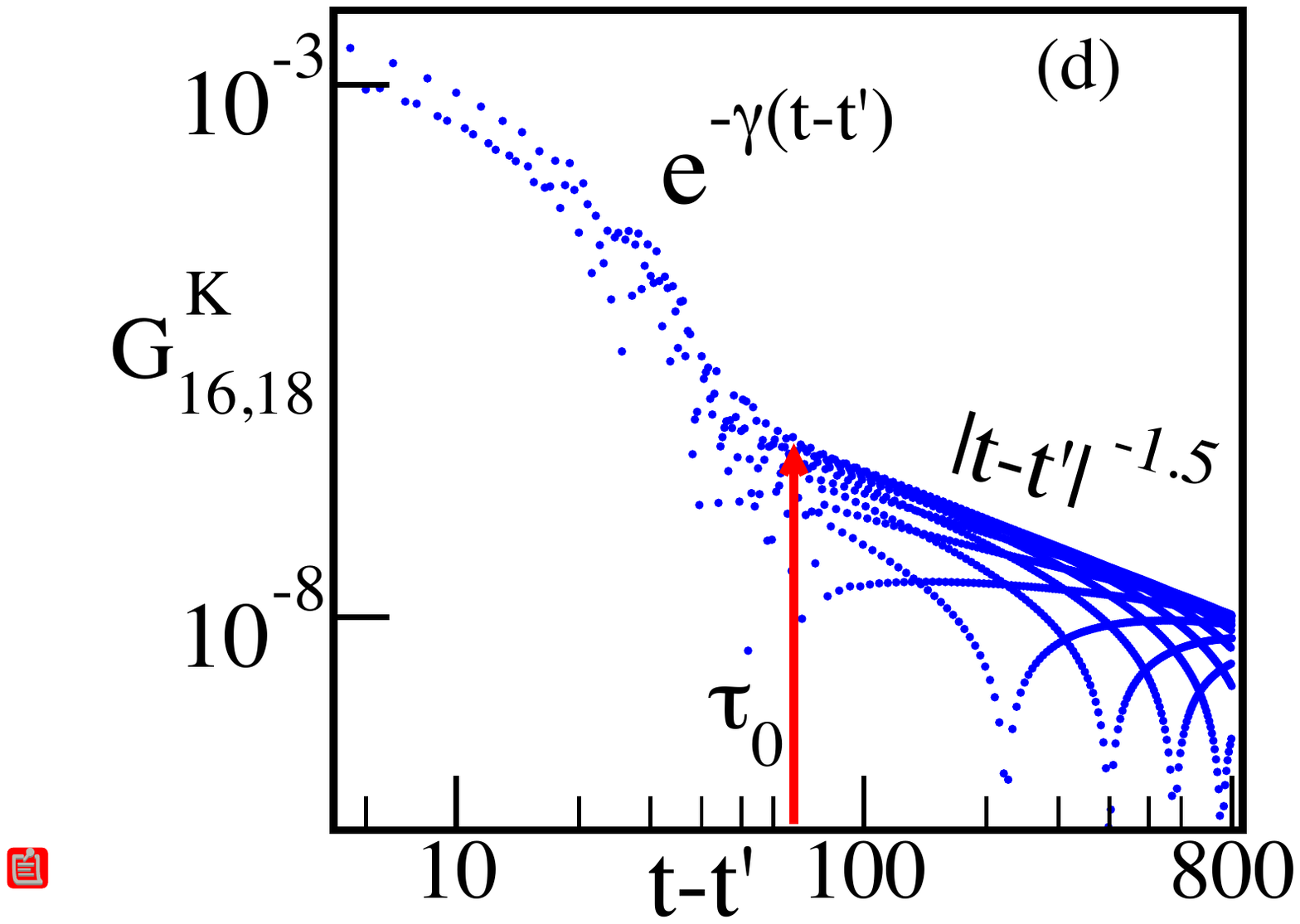}
\caption{(a) and (b): Greens functions of a two-site system connected to two baths with $\mu_1 = -2.5t_B$ and $\mu_2 = -5 t_B$, plotted as function of $t-t'$ for (a) $G^{R}_{1,2}$ and  (b) $Re[ G^{K}_{1,2}]$. (c) and (d): Green's functions for a $N=50$ site chain, where each site is connected to an independent bath, plotted as a function of $t-t'$ in (c) $G^R_{16,18}$ and (d) $Re[G^K_{16,18}]$. The baths have a $\mu$ profile linearly varying with site number, with $\mu_1 = -2.5t_B$ and $\mu_{50} = -5 t_B$. Note that absolute values are plotted in the log-log plot. All the Greens' functions show an initial exponential decay, followed by a power law tail $\sim |t-t'|^{-3/2}$ . The arrows mark the cross-over time scale between exponential decay and the non-Markovian power law tail, $\tau_{0}=\gamma^{-1} \sim\ t_{B}/\epsilon^{2}$. All graphs are calculated for system hopping $g = 0.5 t_{B}$ , system bath coupling $\epsilon = 0.3 t_{B}$ and uniform bath temperature $T=0.625t_B$.}
 \label{figure:greensfn}
 \end{figure}
\section{Analytic Greens' function for a linear chain }\label{sec:green_chain} 

We consider bosons hopping on a 1D chain of $N$ sites where each site is connected to an independent bath with its own temperature and chemical potential. We note that this arrangement is different from the standard transport setup ~\citep{ness_abhishek} , where two reservoirs are connected to the two end sites of the chain. We revert back to eqn.~[\ref{action_diss}] and solve the Dyson equation to obtain the exact retarded and Keldysh Green's functions of this model.

For the 1D chain, the retarded inverse propagator is a symmetric tridiagonal matrix: $G^{-1R}_{ll'}= g [\mathcal{D}~ \delta_{l,l'} +\delta_{l,l'\pm1} ] $, where $\mathcal{D}^{R}(\omega) = g^{-1}[\omega-\Sigma^{R}(\omega)]$ is independent of the site index. This can be inverted ~\citep{analytical_inv} to obtain
\begin{eqnarray}
\label{greens_analytic}
G^{R}_{i,j}(\omega) &=&  (-1)^{i+j} \frac{M_{i-1}M_{N-j}}{g M_{N}} ~~~for~ i<j 
\end{eqnarray}
and $G^R_{i,j}=G^R_{j,i}$ for $i>j$, where 
\beq
M_{i}=\frac{\sinh[ (i+1) \lambda]}{\sinh[\lambda]} ~~ with~~ \cosh \left[ \lambda \right]=\mathcal{D}/2.
\label{M_defn}
\eeq
 The Keldysh Greens' function is then given by,
\begin{equation}
G^{K}_{i,j}(\omega)
=  -\mathbf{i}\epsilon^{2} J(\omega) \sum_{l=1}^{N}~G^{R}_{i,l}(\omega)\coth\left(\frac{\omega-\mu_{l}}{2T_{l}}\right) G^{*R}_{j,l}(\omega).
\label{gkchain}
\end{equation}
To obtain a clear insight about the structure of the Greens' function, we first consider a simplified toy model consisting of 2-sites connected to two different baths ~\citep{Abhishek2016}. In this two site model, 
\begin{equation}
 G^{R}_{\alpha,\alpha}(\omega) = \frac{\mathcal{D}}{g(\mathcal{D}^{2}-1)} ~~and ~~G^{R}_{\alpha,\bar{\alpha}}(\omega)=\frac{-1}{g(\mathcal{D}^{2}-1)}.
\end{equation}
For the bath spectral function given in eqn.(\ref{Jomega_ourbath}), $\Sigma^R(\omega)= \epsilon^2\omega/2t_B^2 - \mathbf{i}~(\epsilon^2/t_B)[1-(\omega+\mathbf{i} \eta )^2/4t_B^2]^{1/2}$. In this case, the Greens' functions $ G^{R}(z)$ have isolated poles at $z_{0}$, where
\beq
z_0 = \frac{g}{1-\frac{\epsilon^2}{t_B^2}} \left[1-\frac{\epsilon^2}{2t_B^2}\right] -\mathbf{i} \frac{\epsilon^2}{t_B\left( 1-\frac{\epsilon^2}{t_B^2} \right) }\left[  1- \frac{g^2}{4t_B^2}  - \frac{\epsilon^2}{t_B^2} \right]^{1/2}.
\label{pole_nint}
\eeq
For $g^2/4+\epsilon^{2}<t_B^2$,  $z_{0}$ has a finite imaginary part leading to an exponential decay in $G^{R}(t,t')$ with a rate $\gamma\sim \epsilon^2/t_B$. For $g^2/4+\epsilon^2>t^2_{B}$ , the pole is on the real axis (outside the bandwidth), leading to an oscillation in the long time limit, similar to the behaviour found by Nori \textit{et. al} in Ref.~[\onlinecite{nori}]. In this paper we will focus on the regime where the presence of the bath leads to damping in the system. Even in this case, there is an additional non-analyticity in $G^{R}(\omega)$ at $\omega=\pm 2t_{B}$, inherited from the non-analytic nature of $\Sigma^{R}(\omega)$. A careful analysis (see Appendix A) shows that the nature of the leading non-analyticity of $G^R(\omega)$ is same as that of $\Sigma^{R}(\omega)$. This leads to a power law tail at long time with $G^R(t-t') \sim (t-t')^{-3/2}$. The Keldysh Green's function also inherits the same power law tail in long time, as can be easily seen from eqn.[\ref{gkchain}]. In Fig.~\ref{figure:greensfn} (a) and (b), we plot respectively the retarded Green's function $G^R_{12}(t-t')$ and the real part of the Keldysh Green's function $G^K_{12}(t-t')$ for the two site model as a function of time in a log-log plot. The plots are obtained for a system with $\epsilon = 0.3 t_{B}$ and $g = 0.5 t_{B}$ connected to two independent baths of common temperature $T=0.625t_B$ and chemical potential $\mu_1 = -2.5t_B$ and $\mu_{2} = -5 t_B$. At short times, the Greens' functions are dominated by the exponential decay from the poles, while the long time behaviour is governed by the power law due to the non-analytic $J(\omega)$. The power law is clearly visible after the exponential part has decayed i.e beyond a time scale $\tau_{0}= \gamma^{-1}\sim \frac{t_{B}}{\epsilon^{2}}$, which is marked in the figures with an arrow. $\tau_0$ is large at weak system-bath coupling and we recover a quasi-Markovian dynamics for a long time, $t<\tau_{0}$ which finally crosses over to a non-Markovian regime for $t>\tau_{0}$. However, at strong system-bath coupling, $\tau _0$ is very short and the dynamics is mostly governed by the non-Markovian power law decay in the memory kernels. Thus, it would be easier to detect observable consequences of non-Markovian dynamics at large $\epsilon$. $\tau_0$ has a very weak dependence on temperature and is essentially set by the system-bath coupling. We emphasis that , in spite of the presence of a scale $\tau_0$ in the system dynamics, we can not coarse - grain over this scale and obtain a Markovian description, as we are then left with the power law decay of the Greens' function.

The characteristic features shown by the two-site system is carried over to the solution for the $N$ site system given in eqn.[\ref{greens_analytic}]. In this case, the poles are given by $\sinh[(N+1)\lambda]=0$, where the location of the poles satisfies $z_{0}(1-\epsilon^{2}/2t_{B}^2)+\mathbf{i}\epsilon^2/t_B \sqrt{1-z_{0}^{2}/4t_{B}^{2}} = 2g \cos [m\pi/(N+1)]$, with $m=1,2, ...N$ ~\citep{pole0}. The pole has a positive imaginary part $\gamma \sim \epsilon^2/t_B$  for $g^2\cos^2[\pi/(N+1)]+\epsilon^{2}<t_B^2$. With this modification, a similar structure of an exponential decay followed by a power law tail is also obtained in this case. We consider a $N=50$ site chain with $\epsilon = 0.3 t_{B}$ and $g = 0.5 t_{B}$, connected to independent baths with common temperature $T=0.625t_B$ and a chemical potential profile which varies linearly with the site number of the system going from $\mu_1 = -2.5t_B$ and $\mu_{50} = -5 t_B$. In Fig.~\ref{figure:greensfn} (c) and (d), we respectively plot the retarded Greens' function $G^R_{i,j}(t-t')$ and the real part of the Keldysh Green's function $G^K_{i,j}(t-t')$ of the system as a function of time in a log-log plot. We have used $i=16$ and $j=18$ to avoid the boundary of the chain. These Green's functions also show a quasi-Markovian exponential decay followed by a non-Markovian $|t-t'|^{-\frac{3}{2}}$ decay, similar to those found in the two site system.

In the next section, we will construct experimentally accessible quantities which can clearly distinguish between the quasi-Markovian and non-Markovian dynamics and discuss the observable consequences of long range memory kernels.

\section{Observables in Steady state}\label{sec:observables}
In this section, we focus on experimental observables which can detect non-Markovian behaviour in these open quantum systems. We will divide these observables into two classes: (i) equal time observables like occupation numbers and currents and (ii) unequal time current-current correlators. While the first class of observables show interesting quantitative deviations from Markovian answers, especially in the limit of large system bath couplings, the unequal time correlators show qualitatively different behaviour indicating the presence of non-Markovian power law tails. 

\subsection{Equal Time Correlators}
 
\begin{figure*}[t]
 \centering
 \includegraphics[width=0.3\textwidth]{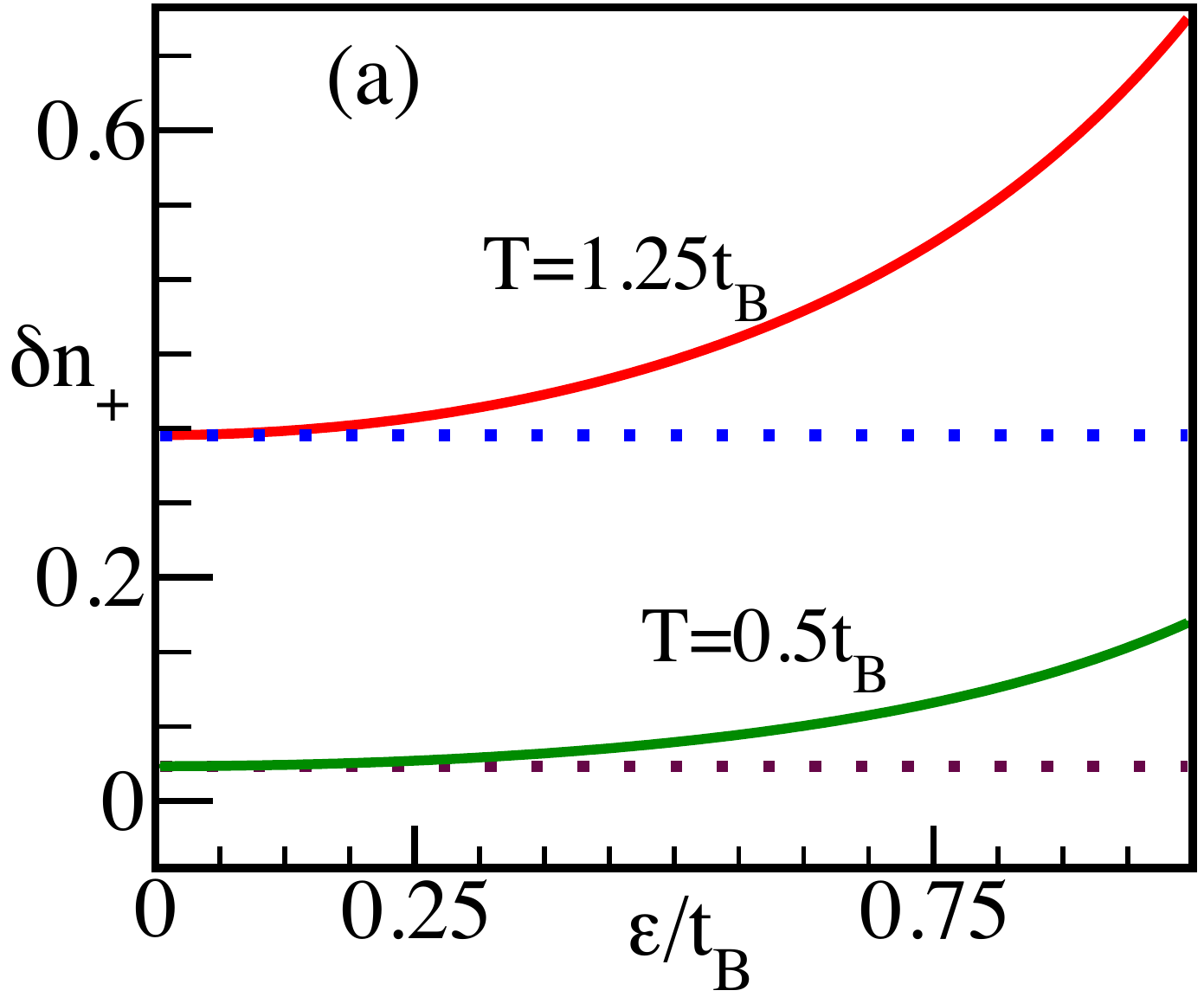}
\includegraphics[width=0.29\textwidth]{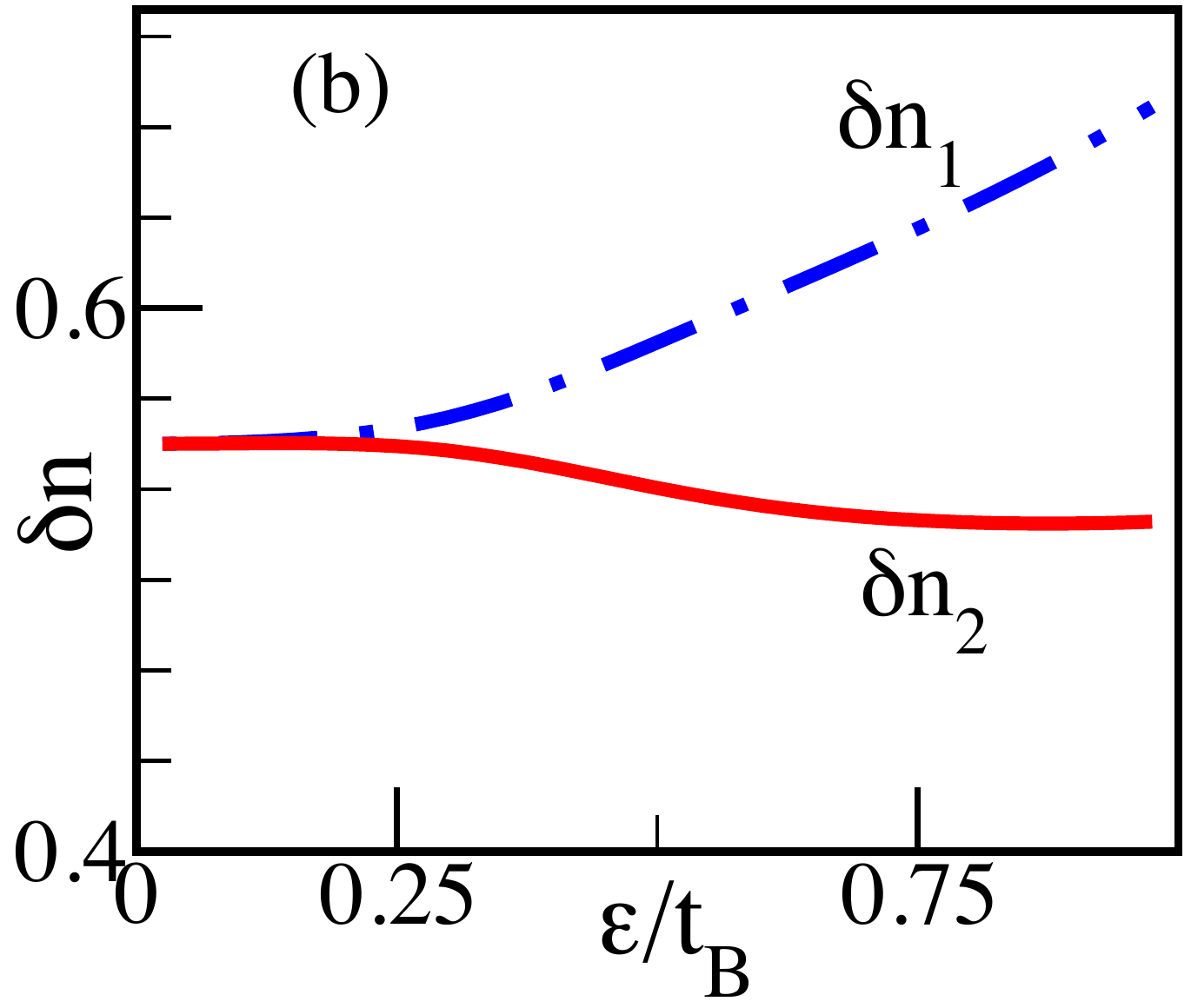} 
\includegraphics[width=0.297\textwidth]{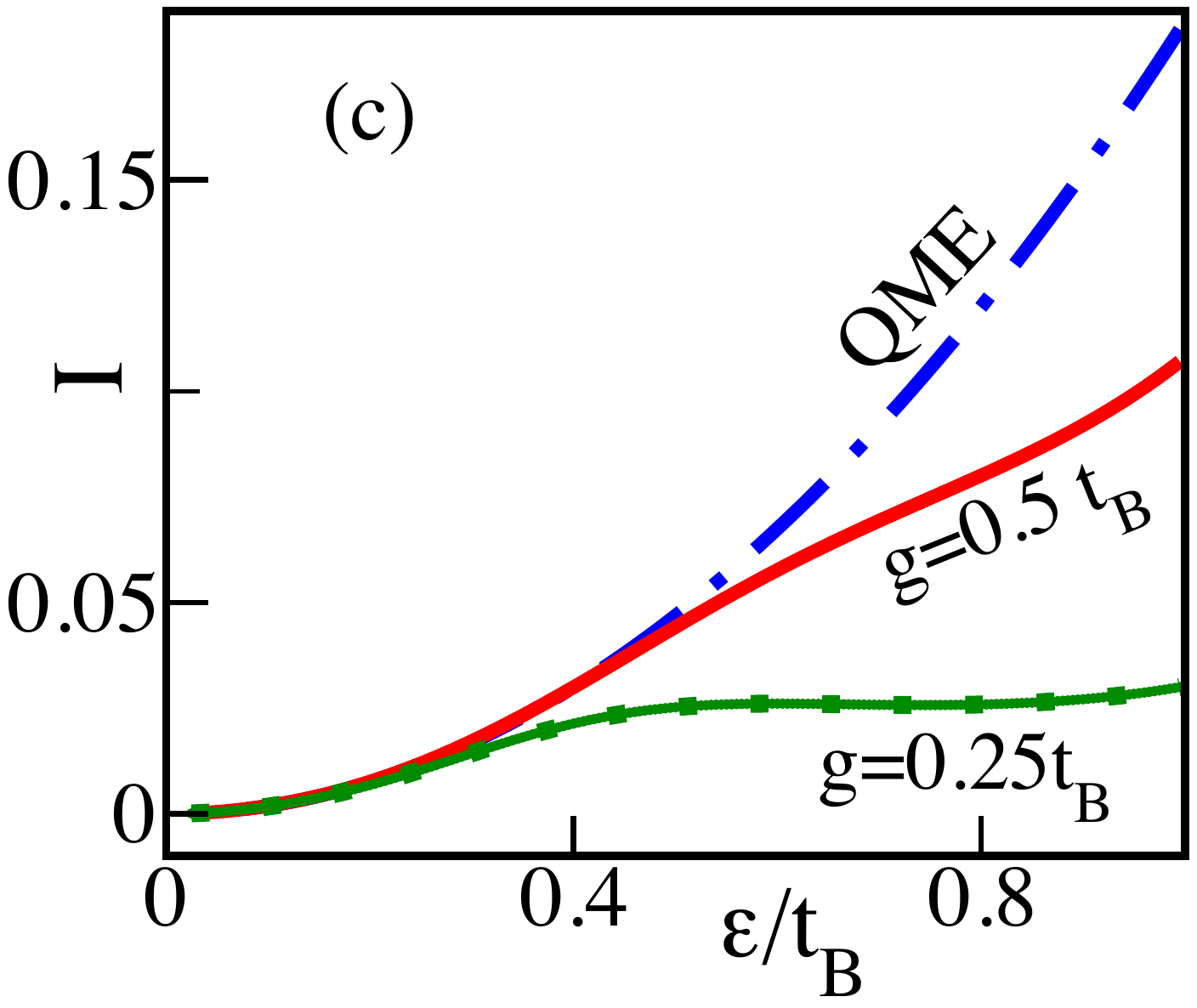} 
\includegraphics[width=0.3\textwidth]{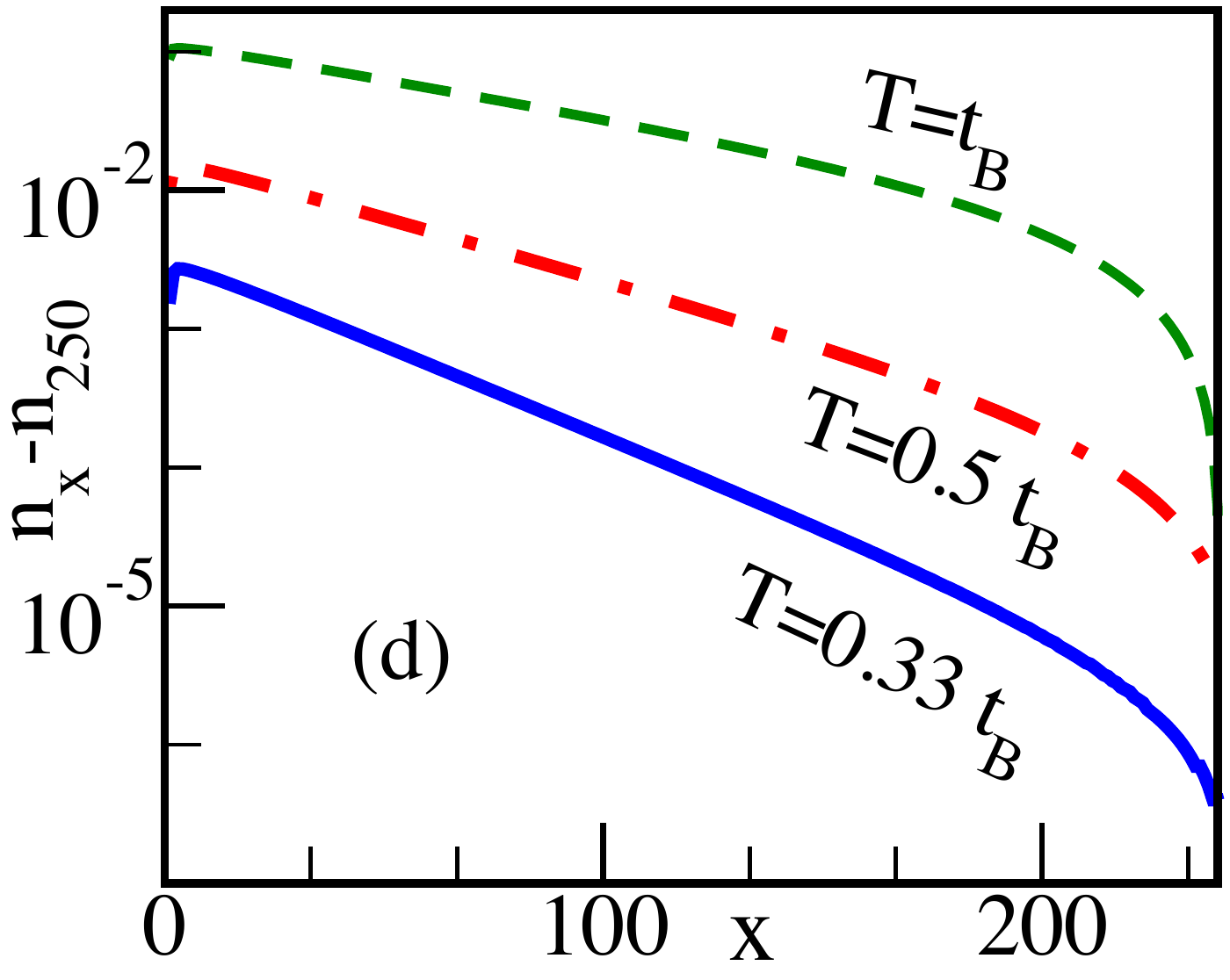} 
\includegraphics[width=0.3\textwidth]{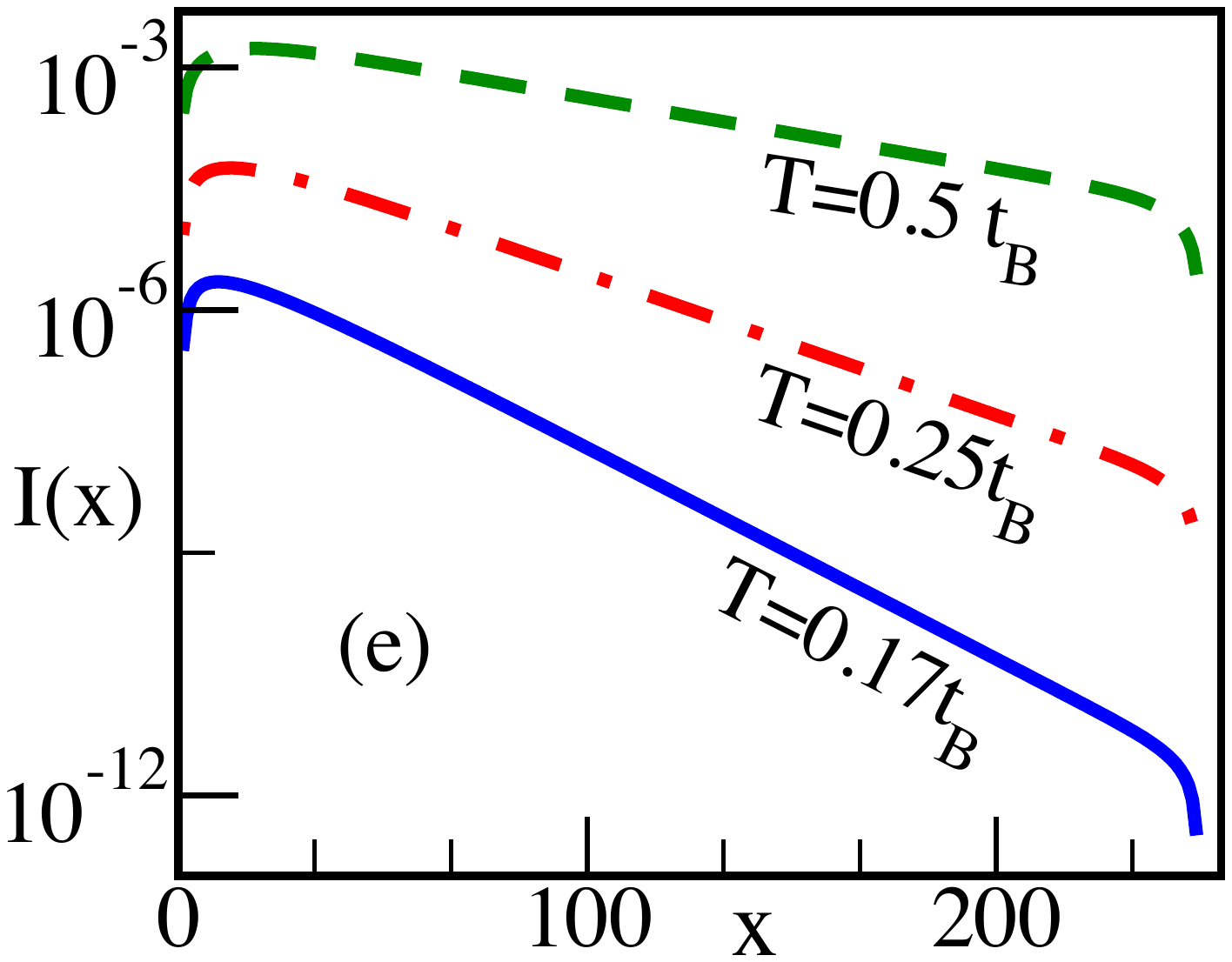} 
\includegraphics[width=0.27\textwidth]{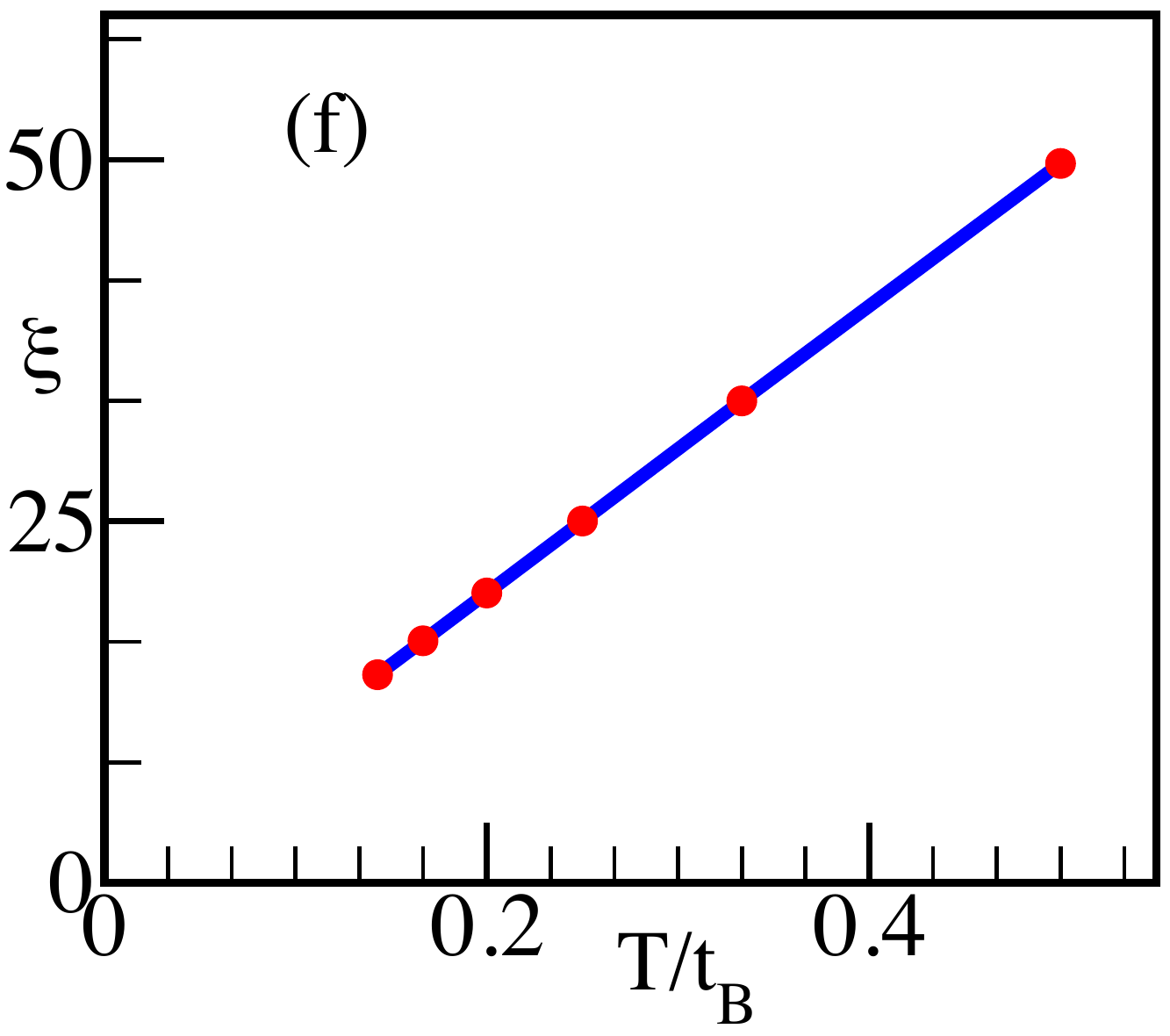} 
\caption{(a) Change in occupation of ground state ($\delta n^+$) of a two site system with $g=0.5t_B$ as a function of system bath coupling $\epsilon$ (solid lines) for coupling to a bath with common $\mu=-2.25t_B$ and common temperature $T=0.5t_B$ and $T=1.25 t_B$. Also plotted are the distribution functions obtained from Markovian master equation given in Ref.[~\onlinecite{Abhishek2016}] (dotted line). The exact answer deviates from the markovian results as $\epsilon$ increases. The deviation increases with temperature. (b)  Occupation number of the two sites and (c) current through a two site system with $g = 0.25t_B$, coupled to two baths with common temperature $T=t_B$ and chemical potentials $\mu_1=-2.5t_B$ and $\mu_2=-5.0t_B$, as a function of $\epsilon$. In (c) the current is also plotted for a system with  $g=0.5t_B $. The steady state current deviates from the quantum master equation result given in Ref.[~\onlinecite{Abhishek2016}] for $\epsilon >g$. (d) Number density profile (measured w.r.t density of $N^{th}$ site) and (e) Current profile for a $N=250$ site chain, where each site is connected to an independent bath. The baths have a $\mu$ profile linearly varying with site number, with $\mu_1 = -2.25t_B$ and $\mu_{250} = -5 t_B$.
In (d) the common temperature for the baths are $T=0.33 t_B, 0.5 t_B, t_B$ while in (e) $T=0.17 t_B, 0.25 t_B, 0.5t_B$. In both cases $\epsilon=0.2 t_B$ and $g=0.5t_B$. Both density and current profiles show an exponential decay. The length scale of the decay $\xi$ is plotted as a function of $T$ in (f). The length scale increases linearly with $T$. }
 \label{figure:observable_equaltime}
 \end{figure*}

We will study two types of equal time correlators: (a) occupation number of sites or eigenmodes and (b) current through the system. We first consider the two site problem. In absence of any coupling to the bath, the system Hamiltonian can be diagonalized to obtain two states at $E=\mp g$ with mode operator $A^{\pm}= (1/\sqrt{2})(a_{1}\pm a_{2})$. The first observable we focus on is the occupation number of these modes, $n^\pm$. The change in occupation number of the modes due to coupling to baths,
\begin{equation}
\delta n^{\pm} = \frac{\mathbf{i}}{4} \sum_\alpha\int\frac{d\omega}{2\pi} \Big[G^{K}_{\alpha,\alpha}(\omega) \pm G^{K}_{\alpha,\bar{\alpha}}(\omega) \Big].
\end{equation}
We first consider the case where a system with $g=0.5t_B$ is coupled to two baths kept at the same temperature and chemical potential. In Fig.~\ref{figure:observable_equaltime}(a) we plot the change in occupation of the ground state, $\delta n^+$, in solid (red and green) lines as a function of the system bath coupling $\epsilon$ for two different bath temperatures, $T=0.5t_B$ and $T= 1.25 t_B $. The common chemical potential of the baths is $\mu=-2.25t_B$. We have also plotted the distribution function obtained from Markovian description given in Ref [\onlinecite{Abhishek2016}], $n^{+}_{eq} = [e^{\frac{- g -\mu}{T}}-1]^{-1}$, in dotted lines. The exact steady state distribution approaches the Markovian answer at small $\epsilon$, but starts deviating as $\epsilon$ grows. This deviation increases with the temperature of the bath. However, the one particle Greens' functions of the two site system do satisfy the fluctuation dissipation relation, $G^{K}_{\alpha,\alpha}(\omega) = 2 \mathbf{i} ~Im\left[ G^{R}_{\alpha,\alpha}(\omega)  \right] ~\coth\left[(\omega-\mu)/2T\right] $, indicative of thermalization of the system. 

We now consider the steady state where the 2 baths coupled to the two sites are kept at same $T$, but different $\mu$. In this case, a finite current flows through the system and it is simpler to analyze the system in the site basis. The change in the local occupation numbers, $\delta n_{1}$ and $\delta n_{2}$, is plotted with $\epsilon$ in Fig.~\ref{figure:observable_equaltime}(b) for $T = t_B$ and $g = 0.25t_B$, $\mu_1=-2.5t_B$ and $\mu_2=-5.0t_B$. As $\epsilon$ increases, the influx of particles to site 1 increases, but the probability of particles tunneling from site 1 to 2 saturates when $\epsilon \geq g$.  The steady state number density on site $1$ then needs to increase to match the outflux with the influx, as seen in Fig.~\ref{figure:observable_equaltime}(b). The density at site 2 simultaneously shows a decrease with $\epsilon$ in the same regime, since the influx to site $2$ has saturated. The number density at site 2 then needs to decrease to to maintain a steady current. Finally, the current, $I_{l } = \mathbf{i} ~g\langle a^{\dagger}_{l}a_{l+1} - a_{l+1}^{\dagger}a_{l} \rangle $, in the link between sites $l$ and $l+1$ is given by,
\begin{equation}
I_l = g\int \frac{d\omega}{2 \pi} ~Re[G^{K}_{l,l+1}(\omega)]
\end{equation}
The current on the link between site 1 and 2 is plotted as a function of $\epsilon$ in Fig.~\ref{figure:observable_equaltime}(c) for two different values of $g = 0.5t_B$ and $ 0.25t_B$ . The current matches with the Born-Markov answer obtained through the solution of Redfield equations~\cite{Abhishek2016} in the small $\epsilon$ limit and deviates from the master equation result for $\epsilon\geq g$. At small $\epsilon$ , the current in the system is constrained by exchange of particles between baths and sites of the system . Any particle that has reached site 1 can be thought to be delocalized to site 2 on a time scale $g^{-1}$ with $g^{-1}<< \epsilon^{-1}$ in the Markovian limit. For $\epsilon \geq g$ , this is no longer true and steady state current is constrained by the hopping rate $g$ and becomes almost independent of $\epsilon$ for $\epsilon >> g$ . This leads to the saturating trends seen in Fig.~\ref{figure:observable_equaltime}(c). The increase of $n_1$ and decrease of $n_2$ in Fig.~\ref{figure:observable_equaltime}(b) is a consequence of this bottleneck between sites $1$ and $2$. 

We now consider a linear chain of $N$ sites where each site $l$ is connected to an independent bath of temperature $T_{l}$ and chemical potential $\mu_{l}$. We focus on the situation where the baths have a fixed temperature $T_{l} = T$ and a chemical potential linearly varying with space i.e $\mu_{l} = \mu_{1} + \nu (l-1)$. Specifically   we look at a $N=250$ site system with $g=0.5t_B$, $\epsilon=0.2 t_B$, $\mu_1 = -2.25t_B$ and $\mu_{250} = -5 t_B$. The density profile in the system shows an exponential decay with site number on top of a constant value, as seen in a semi-log plot in Fig.~\ref{figure:observable_equaltime}(d), where the local density is measured with respect to the density at the $N^{th}$ site. We plot the density profile for three different temperatures, $T=0.33 t_B, 0.5 t_B, t_B$. We find that the decay length scale increases with temperature. The current through the link between sites $l$ and $l+1$, $I_l$, is plotted as function of $l$ in Fig.~\ref{figure:observable_equaltime}(e). The current initially increases with distance from the boundary, then settles into an exponential decay over a large range of sites, as seen in the linear graph in the semi-log plot. We plot the current profile for three different temperatures, $T=0.17 t_B, 0.25 t_B, 0.5t_B$. Once again we find a decay length scale increasing with temperature.
In Fig.~\ref{figure:observable_equaltime}(f), we plot the decay length from the current profile, $\xi$, as a function of temperature and show that $\xi$ is proportional to $T$. In the extreme classical limit, when $T \rightarrow \infty$, the system shows a constant current independent of the link number. The exponential decay and the variation of the decay length with temperature can be understood from the exact Green's function for the linear chain, as shown in Appendix (B). We have checked that a similar exponential decay is also seen when the temperature of the baths vary linearly in space while the chemical potential is kept fixed. 

\subsection{Unequal Time Correlators}
\begin{figure}[t]
\centering
\includegraphics[width=0.238\textwidth]{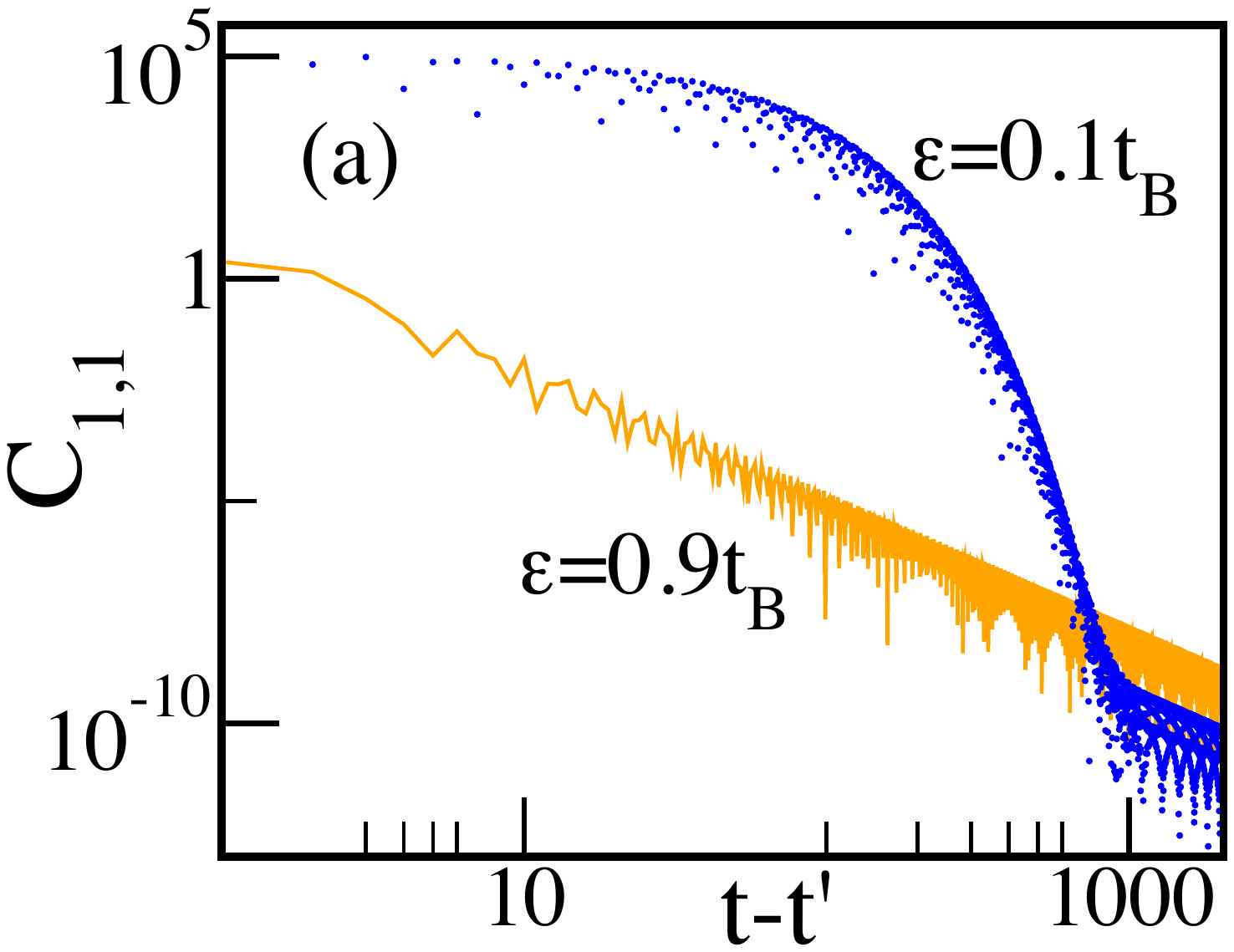}
 \includegraphics[width=0.238\textwidth]{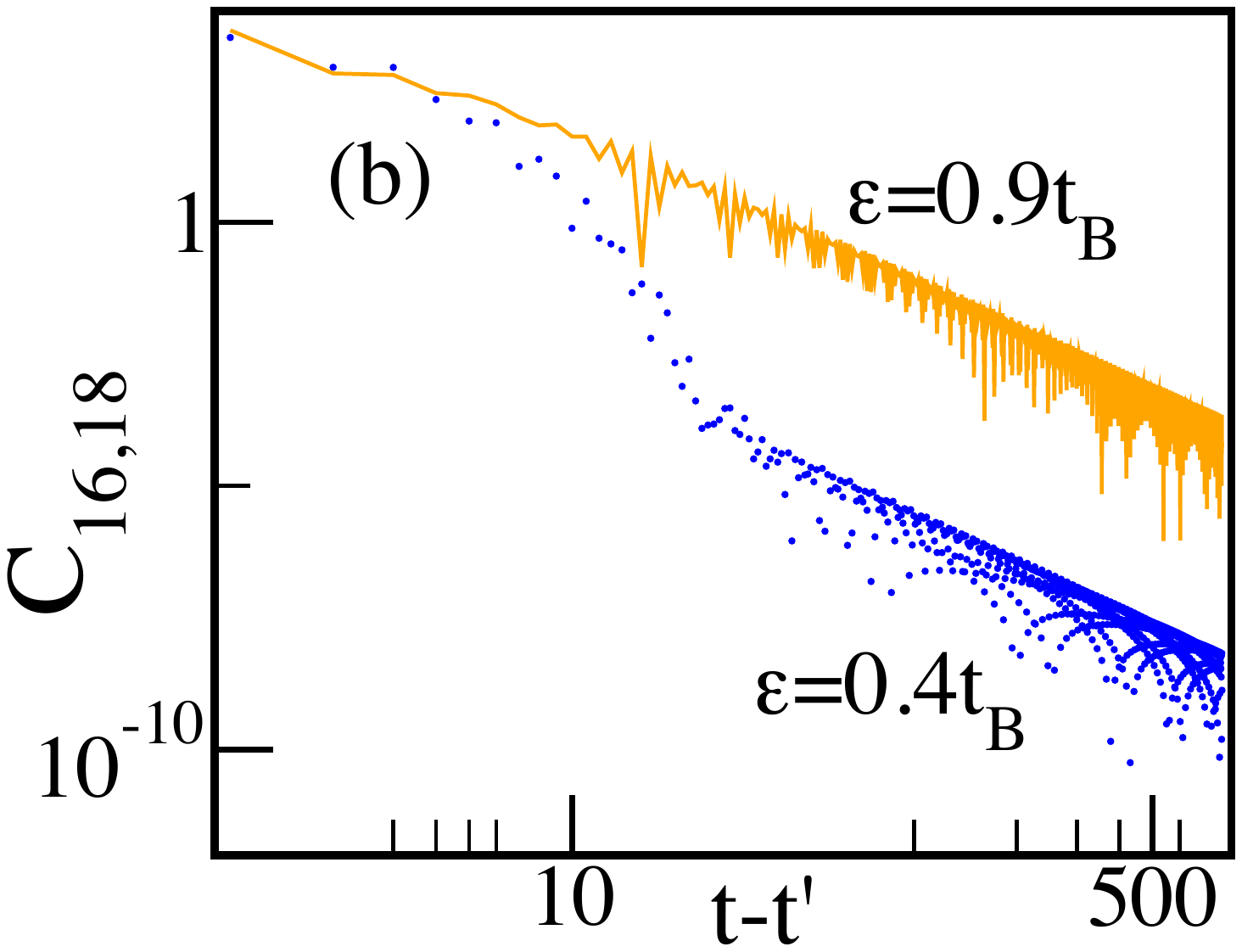} 
  \caption{Unequal time current current correlator $C_{kl}(t-t')$ is plotted as a function of $|t-t'|$ for (a) $N=2$ site model and (b) for a $N=50$ site chain with $k=16,18$. Here each site is connected to a bath. The baths have common temperature $T=0.625t_B$ and linearly varying $\mu_{l}$ where $\mu_{1}=-2.5 t_B$ , $\mu_{N}=-5.0 t_B$ , $g = 0.5t_B$. In (a) the blue dots corresponds to $\epsilon = 0.1t_B$ while the orange line corresponds to $\epsilon = 0.9 t_B$. In (b) the blue dots corresponds to $\epsilon = 0.4t_B$ while the orange line corresponds to $\epsilon = 0.9 t_B$. It shows a short time exponential decay followed by a non-Markovian power law tail $\sim |t-t'|^{-3}$ in the long time limit. The power law tail appears at shorter times as $\epsilon$ increases. }
\label{figure:current_current}
\end{figure}

The density and current profiles in the open quantum system show important quantitative traits as a function of the system bath coupling. However they do not provide a smoking-gun signature of the underlying non-Markovian dynamics. This is provided by the unequal time density-density or current-current correlator, which shows a long time power law decay with an exponent that is twice the exponent of the power law tail in $\Sigma$ and $G$. Here we compute the unequal time current-current correlator , which is given by,
\begin{widetext}
\begin{eqnarray}
C_{kl}(t-t')=\langle I_k(t)I_l(t')\rangle   = g^{2}[G^{<}_{l+1,k}(t',t) ~ G^{>}_{k+1,l}(t,t') +G^{<}_{l,k+1}(t',t) ~ G^{>}_{k,l+1}(t,t') -( l \leftrightarrow l+1)]
\end{eqnarray}
\end{widetext}
where $G^{>(<)}=(G^K\pm (G^R-G^A))/2$. We note that we have considered here a current-current correlator symmetrized between the forward and backward contours; however the qualitative statements we make are also true for a normal ordered correlator. We first focus on the $2$ site system. In Fig~\ref{figure:current_current}(a), we plot $C_{1,1}(t-t')$, normalized by $I_{1}^{2}$, as a function of $t-t'$ on a log-log plot. We plot the correlator for two values of $\epsilon$: a weak system bath coupling of $\epsilon = 0.1t_B$ (blue circles) and a strong system-bath coupling of $\epsilon = 0.9t_B$ (orange line). The graphs are obtained for a system with $g=  0.5t_B$, coupled to two baths with common $T_{1} =  0.625t_B$ and chemical potentials $\mu_{1}= -2.5 t_{B}$ and $\mu_{2} = -5.0 t_{B} $. In the weak coupling limit ($\epsilon=0.1t_{B}$), we see that $C_{1,1}(t-t')$ decreases exponentially in time before crossing over to a power law $C_{1,1}(t-t') \sim |t-t'|^{-3}$ at a very large time $\sim t_{B}/\epsilon^{2}$.  However, by this time the correlator has decayed by orders of magnitude and the power law tail may be hard to observe experimentally. Thus, although the behaviour of the system is not Markovian in the long time limit, experiments may see an effectively Markovian dynamics. However, at large $\epsilon\sim 0.9t_B$, the initial exponential dynamics is extremely short lived and the power law tail in the auto correlation function should be clearly visible in the experiments. This observable can hence be used to identify non-Markovian dynamics in the system and detect the nature of non-analyticity in the corresponding bath spectral function. Further,  one can tune the system dynamics from ``quasi'' Markovian to non-Markovian by tuning the system-bath coupling in these OQS.
 We now consider the current current correlations in a linear chain of $N=50$ sites coupled to independent baths of fixed temperature and linearly varying chemical potential. We ignore the boundary regions (where the current is not exponentially decaying ) and focus on the middle of the chain . In Fig.~\ref{figure:current_current}(b), we plot the current-current correlator $C_{ij}(t-t')$ as a function of $t-t'$ by fixing $i=16$ and $j=18$ for a system with $g = 0.5t_B$ coupled to baths with $\mu_{1} = -2.5t_B$, $\mu_{50}=-5.0t_B$ and $T = 0.625t_B$. The current current correlation function is normalized by $I_{16}I_{18}$ in this case. Once again, we plot $C_{i,j}(t-t')$ for two values of $\epsilon$ : (i) a moderate value of $\epsilon = 0.4t_B$ (blue dots) where an exponential decay is followed by a slow decrease $~|t-t'|^{-3}$ and (ii) a large $\epsilon = 0.9t_B$ where the exponential decay is almost invisible and the power law decay dominates. This regime should be easily observed in experiments.

\subsection{Connection with survival probability}

Non-Markovian dynamics has been traditionally studied by observing non-exponential decay \citep{khalfin,knight_refA} of survival probability of a particular state/density matrix in time. This was first proposed to a closed system, where a discrete mode couples to a semi-infinite continuum by Khalfin et. al \citep{khalfin} and recently extended to provide bounds on decay in presence of baths by Beau et. al \citep{oqs_adelfo}. The detailed calculation of survival probability of a generic density matrix is beyond current scope of Keldysh field theory which can only treat situations in which the system is initially in a thermal state. We can however relate the power law tails in the Greens' functions in our formalism to the power laws seen in the behaviour of survival probability specially for non-interacting system. 
To see this, consider an initial state where a particular mode $\alpha$ has occupation number $n_0$ at initial time $t=0$. The occupation probability of that level at time $t$ is, $n_{\alpha}(t) \sim 0.5~ [(2n_0+1) |G^R(t,0)|^2  + \mathbf{i} \int_0^{t} \int_0^{t} ~dt_1 ~dt_2~G^R(t,t_1) \Sigma^{K}(t_1,t_2) G^{R*}(t,t_2) -1 ] $, where the first term represents propagation which keeps the particle in the excited state, with $|G^R(t,0)|^2$ giving the probability of surviving in this state. Note that this term does not depend on the bath temperature (occupation of bath modes). The second term, which is independent of the initial condition, represents the fluctuations in the occupation due to stochastic exchange with the bath, and hence depends on the temperature of the bath. In a Khalfin \citep{khalfin} like picture, where a semi-infinite continuum is present in a closed system, only the first term would contribute and hence the occupation should decay as $|G^R(t,0)|^2 \sim 1/t^3$ in our specific example. In addition to providing a continuum to scatter from, the bath also provides a stochastic contribution to the occupation number, which should in principle be captured by the extension of Khalfin's work [\onlinecite{oqs_adelfo}]. We can show that the second term leads to a constant occupation in the long time limit, but this constant value is approached as a power law $\sim 1/t^3$. Thus the occupation of the mode will decay from its initial value to a constant value, but $dn_{\alpha}/dt ~\sim 1/t^4$ for large times in this system. We note that $dn_{\alpha}/dt$ is related to the fluorescence intensity measured in Ref [\onlinecite{organic_Monkman}]. Our connection of the power law exponent with the non-analyticites can be used to analyze these experiments.

\section{ Fermionic system coupled to Fermionic baths}\label{sec:fermions}
The formalism developed in the previous sections for treating a bosonic open quantum system with non-Markovian dynamics can be easily modified to treat a system of fermions interacting with fermionic baths. The key features, which make it exactly solvable, are non-interacting Hamiltonians for the system and bath and a linear system-bath coupling. As long as these conditions are met, minor changes in the formalism allow us to treat the fermionic system with similar power law tails obtained for self-energies, Greens' function and unequal time correlators.

\begin{figure}[t]
\centering
\includegraphics[width=0.3\textwidth]{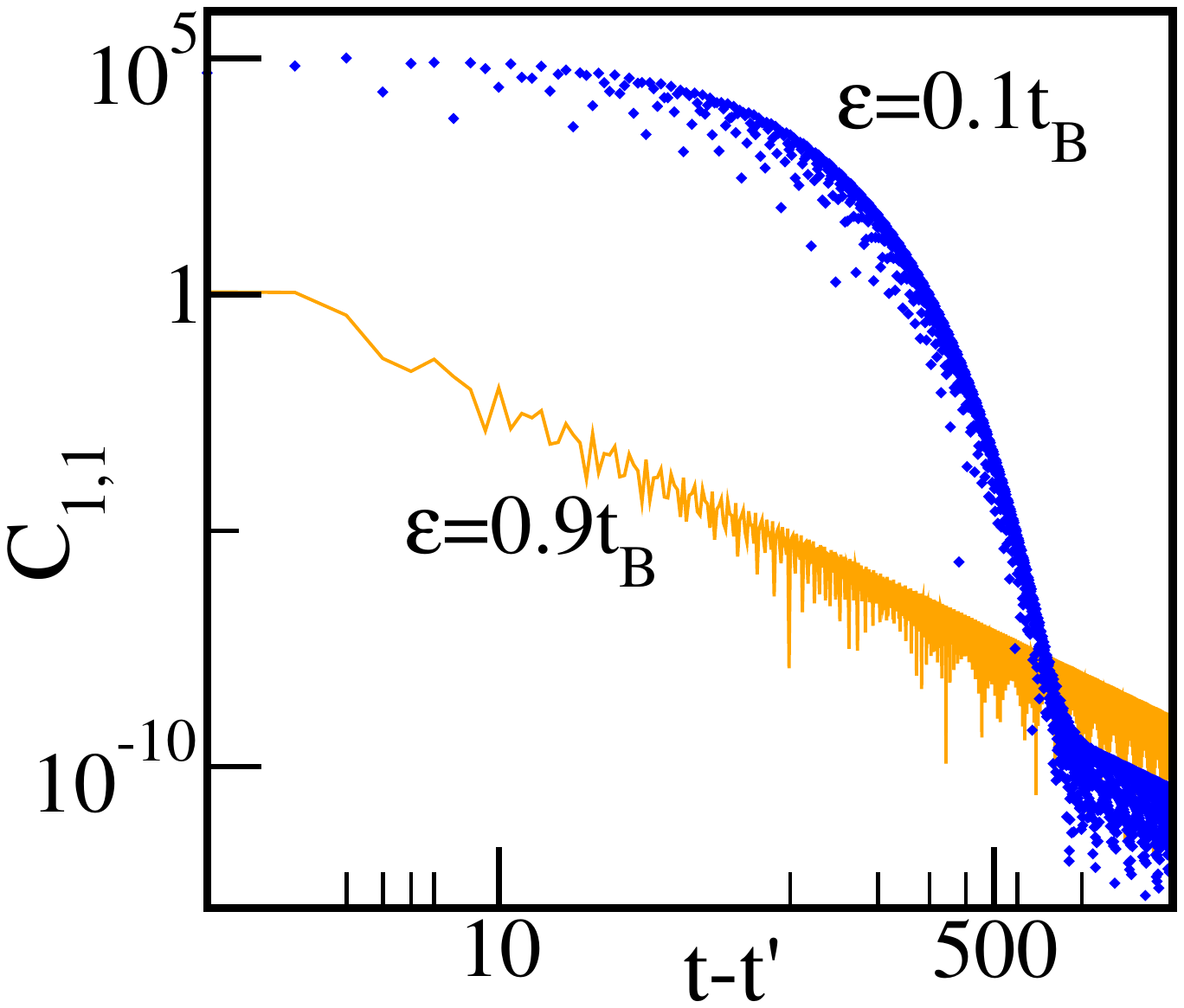}
\caption{The current current correlator $C_{11}(t-t')$ (normalized by $I_1^{2}$) is plotted for a two site fermionic system with $g = 0.5t_B$, coupled linearly to two fermionic baths of same temperature $T=0.625t_B$ and chemical potentials $\mu_{1}=-2.5t_B$ and $\mu_{2}=-5.0t_B$.  It shows a short time exponential decay followed by a non-Markovian power law tail $\sim |t-t'|^{-3}$ in the long time limit. Power law kernel appears at shorter times as $\epsilon$ increases. }
\label{figure:fermion_current_current}
\end{figure}

For fermions, the Keldysh field theory is set up in terms of doubled Grassmann fields $\chi_{\pm}$ and their conjugates $\chi^{*}_{\pm}$. The only difference from the bosonic theory is that it is more convenient to work in terms of $\chi_{1(2)} = [\chi_{+} ~\pm~ \chi_{-}]/\sqrt{2}$ and $\chi_{1(2)}^{*} = [\chi_{+}^{*} ~\mp~ \chi_{-}^{*}]/\sqrt{2}$, which take care of anti-commutation relations between fermionic fields in a natural way. We consider a fermionic chain, where each site is coupled to an independent fermionic bath with its own temperature and chemical potential. The Hamiltonian of the system is given by equations (\ref{model_Ham},\ref{bath_eigen}), with $a^{\dagger}_{l}$ and $B_{\alpha}^{(l)\dagger}$ now representing fermionic creation operators. The Keldysh action for the system is given by,
\begin{equation}
\nonumber
S_{s} = \sum_{l,l'}\int d\omega
\chi^\dagger_{l}(\omega)
    \left[ {\begin{array}{cc}
   G^{-1R}_{0}(l,l',\omega) &G_{0} ^{-1K}(\omega)~\delta_{l,l'} \\ 
   0 & G^{-1A}_{0}(l,l',\omega) 
  \end{array} } \right]
\chi_{l'}(\omega)
  \label{fermion_action_system}
\end{equation}
\begin{equation}
\nonumber
S_{b} = \sum_{l,\alpha}\int d\omega~
\xi^{\dagger(l)}_\alpha(\omega)
    \left[ {\begin{array}{cc}
  \omega-\Omega_{\alpha}+\mathbf{i}\eta &  2\mathbf{i}\eta~F_{l}(\Omega_{\alpha}) \\ 
  0  & \omega-\Omega_{\alpha}-\mathbf{i}\eta
  \end{array} } \right]
\xi^{(l)}_\alpha(\omega)
  \label{fermion_action_bath}
\end{equation}
\begin{equation}
S_{sb} = -\epsilon\sum_{l,\alpha}\int d\omega~ \kappa_{\alpha}\xi^{\dagger (l)}_\alpha(\omega)\chi_l(\omega)+h.c.
 \label{fermion_action_sb}
\end{equation}
where  $\chi^\dagger_l=\left[  \chi^{*(l)}_{1},\chi^{*(l)}_{2}\right]$ are the system fields and  $\xi^{\dagger(l)}_\alpha=\left[  \xi^{*(l)}_{1,\alpha},\xi^{*(l)}_{2,\alpha} \right]$ are the fields for bath degrees of freedom in eigenbasis. The main difference from the bosonic case is the equilibrium distribution function of the fermionic baths $F_{l}(\Omega_{\alpha})=\tanh\Big(\frac{\Omega_{\alpha}-\mu_{l}}{2T_{l}}\Big)$. Integrating out the bath degrees of freedom, we obtain the action for the reduced dynamics of the system,
\begin{widetext}
\begin{equation}
S_{oqs} = \sum_{l,l'}\int d\omega
\chi^{\dagger}_l(\omega)
    \left[ {\begin{array}{cc}
    G^{-1R}_{0}(l,l',\omega)-\Sigma^{R}(\omega) ~\delta_{l,l'}& -\Sigma^{K}_{l}(\omega)~\delta_{l,l'}\\ 
  0 & G^{-1A}_{0}(l,l',\omega)-\Sigma^{A}(\omega)~\delta_{l,l'} \\
  \end{array} } \right]\chi_{l'}(\omega)
  \label{fermion_action_diss}
\end{equation}
\end{widetext}
where $\Sigma^{R}(\omega) $ is given by equation [\ref{selfenergy:freq}], just as in case of bosons, while the Keldysh self energy $\Sigma^{K}_{l}(\omega)$ is given by,
\begin{equation}
\Sigma^{K}_{l}(\omega)  =-\mathbf{i}\epsilon^{2}J(\omega) ~ \tanh\Big(\frac{\omega-\mu_{l}}{2T_{l}}\Big)    
\end{equation}
The Keldysh self energy function for the fermionic case can thus be obtained from the bosonic case by replacing $\coth\Big(\frac{\omega-\mu_{l}}{2T_{l}}\Big) $ by $\tanh\Big(\frac{\omega-\mu_{l}}{2T_{l}}\Big) $, as one would expect due to different statistics of particles. 

With this replacement, all the expression for the Greens' functions and observables, obtained in section  \ref{sec:diss_noise} , \ref{sec:green_chain} and \ref{sec:observables} for bosonic system, can be used for the fermionic system as well. We note that, unlike the bosonic case, there is no reason to place the bath chemical potentials below the band bottom for the fermionic baths. The self-energies in real time show similar power law behaviour $\Sigma^{R}(t-t') \sim \Sigma^{K}_{l}(t-t') \sim |t-t'|^{-\frac{3}{2}}$ for large time, with the power law originating from the non-analyticities in the bath spectral function given by equation [\ref{Jomega_ourbath}]. However, it is harder in this case to give an interpretation to self-energies, since fermions do not have a classical limit and it is impossible to talk about a \textquotedblleft classical saddle point" for non-interacting fermionic action. We note that the mere presence of the singularity in the bath spectral function is enough to generate the power law tail, the singularity does not need to be either close to the spectral gap of the system or to the chemical potential of the bath. The Greens' functions then inherit this power law along with an exponential decay, which also gets reflected in the long time behaviour of density-density auto-correlation function and current-current correlation function, similar to the bosonic system considered earlier. Working with the 2-site fermionic system, we plot $C_{1,1}(t-t')=\langle I_1(t)I_1(t')\rangle$, normalized by $I_{1}^{2}$, as a function of $t-t'$ on a log-log plot in Fig.~\ref{figure:fermion_current_current} for a system with $g = 0.5t_B$ coupled to baths with $\mu_{1} = -2.5t_B$, $\mu_{2} = -5.0t_B$ and $T = 0.625t_B$. The current-current correlator clearly shows an exponential decay followed by a power law tail $\sim |t-t'|^{-3}$, as seen in the bosonic system. Once again the power law dominates the dynamics and leads to large values of the correlators for long $t-t'$ at strong system bath coupling.

 We have obtained exact solutions for Green's functions of 1D non-interacting bosonic and fermionic chains coupled to independent baths and demonstrated non-Markovian behaviour in both systems.  An obvious question is how the interaction among system degrees of freedom, which is inadvertently present in any realistic system, affects the description we have obtained here. In the next section, we try to answer this question by focusing on the bosonic system and making mean field approximation to the underlying interaction terms. 
\section{Effect of Interaction}\label{sec:interactions}
\begin{figure}[h]
\centering
\includegraphics[scale=0.35,valign=t]{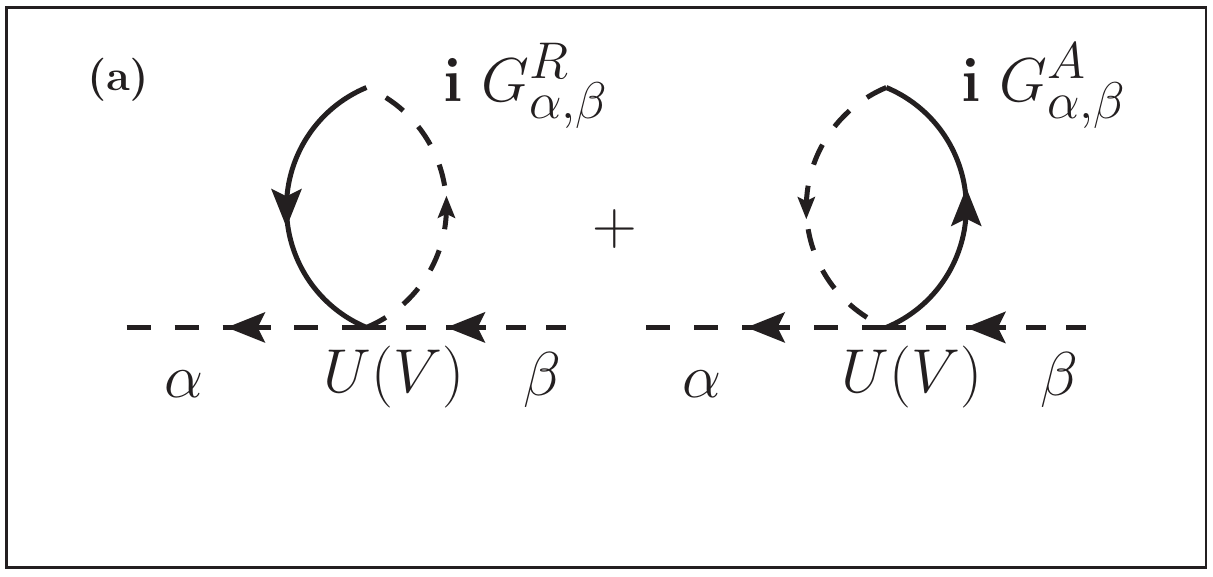}  
\includegraphics[scale=0.19,valign=t]{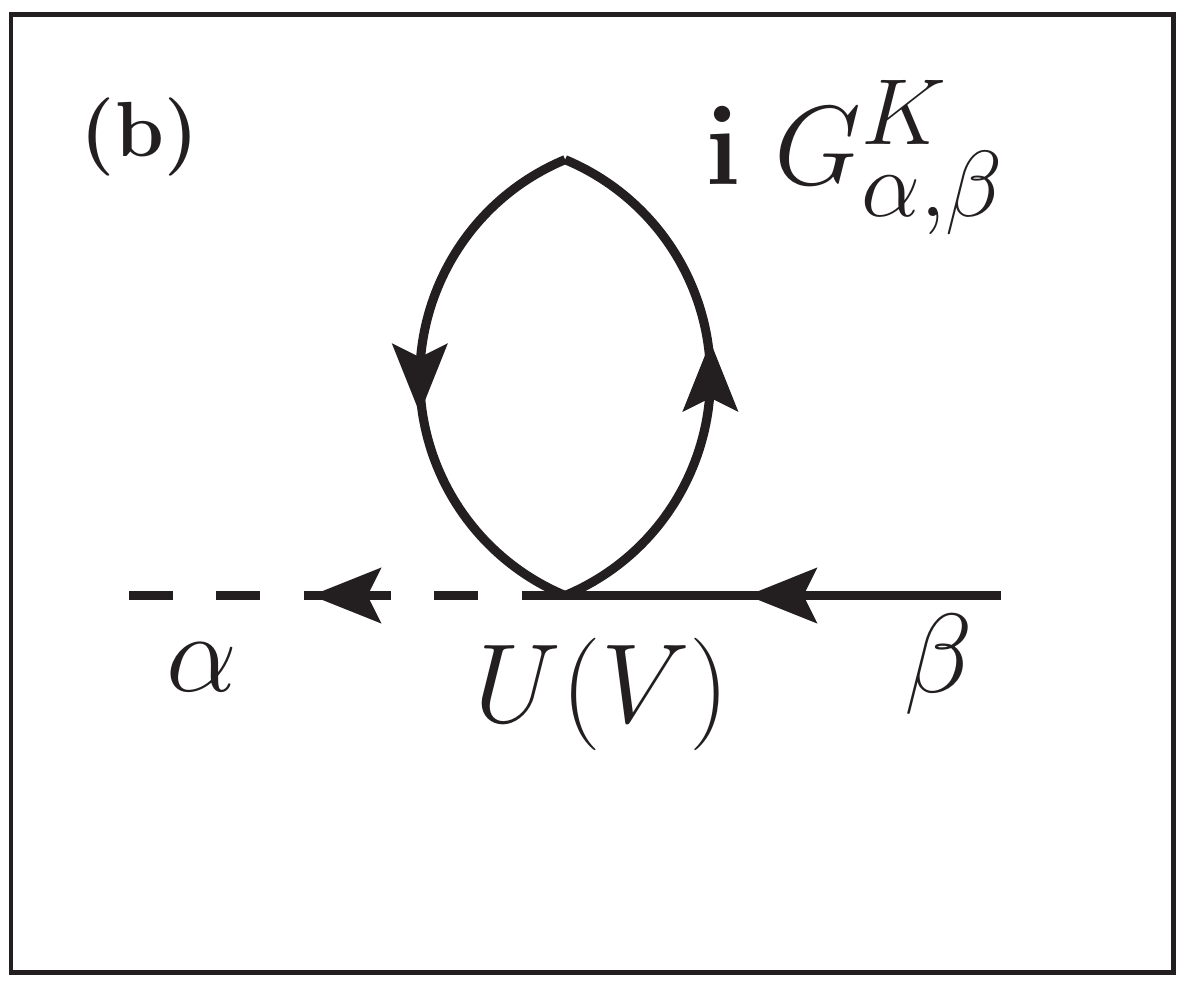} 
\includegraphics[scale=0.18,valign=t]{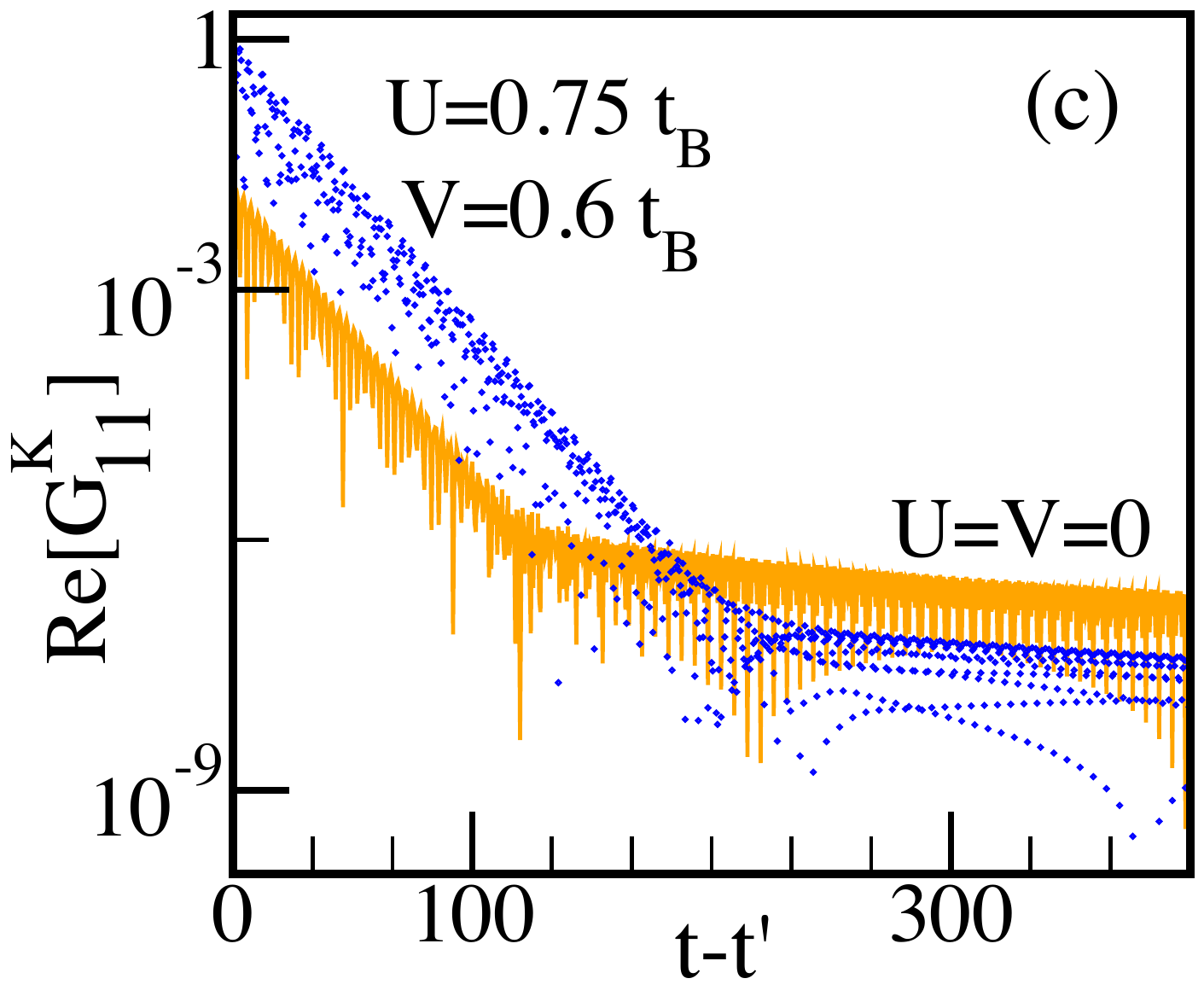}
\caption{(a) and (b): Feynman diagrams corresponding to (a) Keldysh and (b) Retarded self energy due to interparticle interactions in a two site bosonic system coupled to external bath in the mean field approximation. The correction to Keldysh self energy vanishes. The loop propagator in (b) is the self consistent Keldysh Green's function. (c) The real part of the local Keldysh Green's function $G^K_{11}(t,t')$ for a two site bosonic system coupled to two baths plotted as a function of $t-t'$ in a semi-log plot. In both cases the bath temperature $T=0.625t_B$ and chemical potentials $\mu_{1}=-2.5t_B$ and $\mu_{2}=-5.0t_B$. The systems have a hopping $g = 0.5t_B$ and system-bath coupling $\epsilon=0.2t_B$. The orange line is for a non-interacting system, while the blue circles are for an interacting system with $U=0.75 t_B$ and $V=0.6 t_B$. The interaction increases the crossover scale, but does not eliminate the power law tail. }
\label{figure:gk11_interaction}
\end{figure}

We consider the case of an interacting chain of bosons coupled linearly to independent non-interacting bosonic baths. In addition to the hopping $g$, the system bosons are interacting with each other with a local repulsion $U$  and nearest neighbour repulsion $V$, i.e. We add to the Hamiltonian given by equation [\ref{model_Ham}], a term
\begin{equation}
H_{int} = U \sum_{i} \hat{n_{i}} (\hat{n}_{i}-1) + V \sum_{i} \hat{n}_{i} \hat{n}_{i+1},
\end{equation}
 where $\hat{n}_{i}= a^{\dagger}_{i}a_{i}$ is the density at site $i$. The bath degrees of freedom can be integrated out as before to obtain the quadratic dissipative action $S_{oqs}$ ( eqn.[\ref{action_diss}]). This action and its associated Green's functions can then be used as a free theory around which we consider the effects of interactions. The Keldysh action corresponding to the inter-particle interaction is given by~\cite{kamenev},
 \begin{widetext}
\begin{eqnarray}
S_{int} = -\frac{U}{2}\sum_{l} \int dt \phi _{cl}^{*(l)}(t)\phi^{(l)}_{cl}(t)  \phi^{*(l)}_{cl}(t) \phi^{(l)}_{q}(t)  
-\frac{V}{2}\sum_{l} \int dt \phi^{*(l)}_{cl}(t)\phi^{(l)}_{cl}(t)  \phi^{*(l\pm 1)}_{cl}(t) \phi^{(l\pm 1)}_{q}(t) +  h.c  +cl \leftrightarrow q.
\end{eqnarray} 
 \end{widetext}
 We consider the effect of this interaction on the Greens' functions within mean field theory. This is equivalent to considering the one-loop corrections to the self energy, shown in Fig.~\ref{figure:gk11_interaction} (a) and (b) for the Keldysh and retarded self-energies. Note that the Keldysh self energy correction vanishes because $G^R(t,t)+G^A(t,t)=0$ ~\citep{kamenevbook}, while the retarded self-energy due to the interaction is purely real. This is related to the fact that the one loop terms do not lead to redistribution of energy among the modes. The corrections to self-energy, $\Sigma^{R}$ induced by the interaction term are given by, $\Sigma^{I}_{ii} = Un_{i,i}+(V/2)n_{i\pm 1,i\pm 1}-U$, and $\Sigma^{I}_{i,i+1} = (V/2)(n_{i,i+1})$, where $n_{ij}=\mathbf{i}\int (d\omega/2\pi)~  G^{K}_{i,j}(\omega)$. The retarded self-energy due to inter-particle interaction change the retarded Greens' functions obtained from $(G^{-1R}_{l,l'} - \Sigma^{I}_{l,l'})^{-1}$ where $G^{-1R}$ already includes self-energies induced by the bath. These retarded Greens' functions in turn change $G^{K}$ through $G^{K} = G^{R}\Sigma^{K} G^{A}$ where $\Sigma^{K}$ is the Keldysh self-energy induced by the bath (the interaction contribution to it is zero, as shown above). The Keldysh Greens' function used in calculating the retarded self-energy shown in Fig. \ref{figure:gk11_interaction} (b) is this renormalized $G^{K}$, obtained self-consistently. Hence, this completes a self-consistency loop. The mean field approximation is then equivalent to resummation of particular set of loop diagrams. From an equation of motion perspective, this is also equivalent to solving stochastic Gross-Pitaevskii equation~\cite{stoch_Gross}.

 A key question we want to answer is how does the interaction affect the non-Markovian dynamics? We have seen earlier that the Greens' functions and unequal time observables for non-interacting systems show a pattern of exponential decay at short times, which crosses over to power law tail with exponent fixed by the nature of the non-analyticity in the bath spectral function. We can then ask if the power law tail survives in the interacting system, and, if it survives, whether the crossover time scale increases or decreases with the interaction strength. 

In order to gain analytic insight into the problem, we once again consider the case of a chain with 2-sites. In this case, $\Sigma^{I}_{11}=U n_{11} + (V/2) n_{22}-U$ , $\Sigma^{I}_{22}=U n_{22} + (V/2) n_{11}-U$ and $\Sigma^{I}_{12}=(V/2) n_{12}$. The retarded Greens' function is then given by
 \begin{equation}
 G^{R}(\omega)= \frac{1}{Q(\omega)}\left[ \begin{array}{cc}%
   \omega - \Sigma^{R} -\Sigma^{I}_{22} & -g+\Sigma^{I}_{12}\\ 
  -g+\Sigma^{*I}_{12}& \omega - \Sigma^{R} -\Sigma^{I}_{11}
  \end{array} \right] 
  \label{gr_interacting}
 \end{equation}
where $Q(\omega)=[\omega - \Sigma^{R} -\Sigma^{I}_{11}][\omega - \Sigma^{R}-\Sigma^{I}_{22}]-|\Sigma^{I}_{12}-g|^2$.
 Considering the denominator of the equation [\ref{gr_interacting}], we find that the poles of the Greens' function are shifted from $z_0$ in the non-interacting case to $z_\pm$, where $z_\pm$ is obtained from $z_0$ (given in eqn.[\ref{pole_nint}]) by replacing $g$ with $c_{\pm}=a\pm \sqrt{b^2+|\tilde{g}|^2}$. Here $a(b) = (\Sigma^{I}_{11} \pm \Sigma^{I}_{22} )/2$,  and $\tilde{g} = -g + \Sigma^{I}_{12} $. The interaction induced self-energy splits the single pole of the non-interacting case into two. 
Using the fact that $a= (U+V/2)(n_1+n_2) +V/2$, where $n_{1(2)}$ are the densities at site $1(2)$, one can show that $a>0$, and hence $c_+ >g$, for small $\Sigma_{12}$. We have checked that $\Sigma^{I}_{12}$ is not large enough to change this argument for a wide range of $U$ and $V$ (see Appendix C for details). Comparing $z_+$ and $z_0$, we then see that $z_+$ has a smaller imaginary part compared to $z_0$. This leads to an exponential decay slower than the non-interacting case. The power law tail survives (in fact it can be shown to survive to all orders in perturbation theory), but the crossover scale is further pushed out because of the slower exponential decay. In Fig.~\ref{figure:gk11_interaction}(c), we plot the real part of the Keldysh Greens' function $G^{K}_{1,1}$ of the two site system with the same non-interacting parameters $g =  0.5t_B $, $\epsilon= 0.2t_B$, bath temperature $T=  0.625t_B$, $\mu_{1}= -2.5t_B$, $\mu_{2}= -5.0t_B $ for two cases in a semi log plot : (i) the non-interacting case plotted with orange lines and (ii) the interacting system plotted with blue circles for $U=0.75t_B$ and $V =0.6t_B $. We clearly see that crossover to the power law tail is pushed out to larger values of $|t-t'|$ for the interacting case. In Appendix C, we have shown how the crossover timescale changes with $U$ and $V$ for a range of interaction strengths.
 
The fact that the exponential decay is slowed down in presence of the interaction seems odd, given our expectation of faster decay due to scattering events. This is due to fact that the mean field theory misses processes leading to redistribution of energy, which contribute at two loops or higher order. The one loop corrections lead to the dressing of the $1$ particle Greens' function (to form quasi particles) without leading to additional energy relaxations.  
Thus, although the mean field theory is non-perturbative in interaction strength, we expect that at larger interaction strengths, additional process may increase the rate of exponential decay of the Greens' function, thus pulling back the timescale for crossover from ``quasi'' Markovian to non-Markovian dynamics in the system. 
 
\section{Conclusions}
We have used a Schwinger Keldysh field theory to describe the dynamics of a bosonic (fermionic) system coupled linearly to a non-interacting bosonic (fermionic) bath. Integrating out the bath degrees of freedom we obtain the effective dissipative action for the open quantum system. We identify the retarded and Keldysh self energies induced by the bath with the dissipative and noise kernel in a non-local stochastic Schrodinger equation. We show that presence of non-analyticities in the bath spectral function leads to power law tails in these kernels for both bosonic and fermionic system. The exponent of the power law is governed solely by the nature of the non-analyticity and is independent of the location of the non-analyticity. This leads to retarded and Keldysh Green's functions showing an exponential decay followed by a power law tail with the same exponent as the kernels. The crossover timescale is set by the system bath coupling and decreases quadratically with increasing system bath couplings. It is independent of the temperature and chemical potential of the baths. Thus the non-Markovian dynamics, characterized by the power law tails, are easier to observe in systems with stronger system bath couplings. We note that for bosonic baths with chemical potentials, the spectrum must be bounded from below, ensuring at least one point of non-analyticity, and hence non-Markovian dynamics will be ubiquitous in such systems. The power law tails in the Green's functions lead to corresponding power laws in unequal time correlators like the current-current correlator, which can be measured by noise spectroscopy. 

As a concrete example, we consider a linear chain of bosons (fermions) hopping between nearest neighbours, where each site is coupled to an independent bath of non-interacting bosons (fermions). The bath consists of another 1-dimensional lattice of bosons (fermions) with nearest neighbour hopping. We consider a linear coupling between the system and the bath, where the system is coupled to an edge site of the bath. This leads to a bath spectral function which has square root derivative singularities at the band edge of the bath, leading to a $(t-t')^{-3/2}$ power law in the dissipative and noise kernels. We obtain analytic expressions for the self energies and hence for the exact Green's functions of the open quantum system of 1-d chain, and verify the existence of the power law tails. We use a two site model, with simpler expressions for Green's functions to illustrate the main features, before showing that the features are retained in the exact Green's functions for the full 1-d chain.

We focus our attention on two classes of observables in the open quantum system: equal time correlators like density and current, and unequal time correlators like current-current correlation functions. We first consider a two site system coupled to two baths of same temperature and chemical potential and show that the steady state mode occupation numbers deviate from the thermal equilibrium answers with increasing strength of system bath couplings which cannot be explained by a simple dressing of the system spectrum. For the same system, if the two baths are kept at different chemical potentials, as the system-bath coupling becomes larger than the hopping within the system, both local densities and current through the system deviates from the answers provided by the quantum master equation approach. The densities show a pileup on one site compared to the other, while the current saturates for $\epsilon >g$. We then consider the 1-d chain, with sites coupled to independent baths at the same temperature, but with a chemical potential profile. For a linear variation of $\mu$ with space, we find an exponential decay of the current with distance. The density also shows an exponential variation in space on the top of a constant value. The length scale obtained from the decay of the current increases with temperature, with a spatially independent current obtained in the limit of infinite temperature. This behaviour of the current is understood analytically from the exact Green's function obtained for the chain.

We next focus our attention on the unequal time current-current correlators and show that they follow a pattern similar to the Green's functions: a short time exponential decay, followed by a power law tail $\sim (t-t')^{-3}$. This occurs both in the two site system and in the linear chain, and is common to both bosonic and fermionic systems. The timescale for the crossover from exponential to power law decay decreases while the value of the correlator in the power law regime increases with system-bath coupling. These power law tails can then be used to detect non-analyticities in the bath spectral functions, and hence possible phase transitions in the baths.

We finally consider the effect of interactions on the above picture. The power law tails survive to all orders in perturbation theory in the interaction strength. Within a mean field theory for the bosonic two site system, which is equivalent to solving a stochastic Gross Pitaevski equations, we find that the exponential decay slows down and the crossover timescale for observing the power law tails is pushed to larger values. We understand this analytically in terms of the structure of the Green's functions in the system. Increasing interaction strength, should, in principle lead to a larger scattering rate and hence to faster decays, but such two loop processes redistributing energy is not captured within the mean field theory. This needs to be further investigated, although our prediction of a larger crossover timescale should hold in the weakly interacting interacting system.

 \begin{acknowledgments}
The authors thank Abhishek Dhar, Manas Kulkarni, Barry Garraway, Takis Kontos, Bijay Kumar Agarwalla and Archak Purakayastha for useful discussions during the workshop Open Quantum System at ICTS,India. Authors specially thank Carlos Bolech for his suggestions on the effect of interaction. The authors also acknowledge computational facilities at Department of Theoretical Physics, TIFR Mumbai. 
\end{acknowledgments}

\appendix

\section{ Power Law tails and non analytic bath Spectral Functions} 
In section (\ref{sec:diss_noise}) and (\ref{sec:green_chain}) of the main text, we have claimed that the non-analyticity of the bath spectral function leads to power law tails in self energies and Greens' functions using Keldysh formalism. We have also claimed that the exponent of the power law depends solely on the nature of the non-analyticity and is independent of its location. In the main text we had shown this with an exact expression for retarded self-energy for the spectral function of semi-infinite bath having $J(\omega) = \Theta(4t_B^2-\omega^2)(2/t_{B}) \left( 1-\omega^2/(4t^2_{B}) \right)^{1/2}$. 
In this appendix , we make the connection between non-analyticites of $J(\omega)$ and power law tails clearer by showing the relation between the non-analyticity and the exponent of the power law for $\Sigma^{K}(t-t')$ and $G^{R/K}(t-t')$ for the semi-infinite bath, where the Fourier transform cannot be obtained in a closed form. To do this, we will first focus on non-analyticity in $\Sigma^{K}(\omega)$ for $J(\omega)=\Theta(4t_B^2-\omega^2)\frac{2}{t_{B}} \sqrt{1-\frac{\omega^2}{4t^2_{B}}}$ and show how this is connected to power law exponent . This connection will help up in understanding the nature of power law for any non-analytic spectral function , if we know its leading order non-analytic piece. We will then use this argument to show that $\Sigma^{K}(\omega) , G^{R}(\omega),G^{K}(\omega)$ all have the same leading singularity and hence the same power law tail appears in their long time profile. The Keldysh self energy is given by,
\begin{equation}
\Sigma^K_l(t-t') =-\mathbf{i}\epsilon^2 \int \frac{d\omega}{2\pi} J(\omega)\coth \left[\frac{\omega-\mu_l}{2T_l}\right] e^{-\mathbf{i}\omega(t-t')}
\label{supp:sigK}
\end{equation}
We note that the chemical potential, $\mu_{l}< -2t_B$ and hence $\coth \left[\frac{\omega-\mu_l}{2T_l}\right] $ is a smooth function in the integration range and $J(\omega)$ and hence $\Sigma^{K}(\omega)$ is non-analytic only at $\omega = \pm 2t_B$ . Expanding $w=\pm 2t_B(1-\delta)$ we get $J(\omega) \sim \delta^{\frac{1}{2}}$ for $\delta \rightarrow 0^{+} $ and hence the integrand $\mathcal{I}(\omega) =\frac{1}{2\pi} J(\omega)\coth \left[\frac{\omega-\mu_l}{2T_l}\right]$ can be expanded as $\mathcal{I}(\pm2t_B(1-\delta)) =\sum_{n=0}^{\infty} C^{\pm}_{n}~\delta^{n+\frac{1}{2}}$ where $C^{\pm}_0 = \frac{\sqrt{2}}{\pi~t_B} \coth \left[\frac{\pm 2t_B-\mu_l}{2T_l}\right]$. Considering the combination of the non-analytic piece , we get,
\begin{widetext}
\begin{eqnarray}
\Sigma^K(t-t') &=& -\mathbf{i}\epsilon^{2}~2t_{B}\sum_{n=0}^{\infty}\Big[ \int_{0}^{1} d\delta ~ C^{+}_{n}~\delta^{n+\frac{1}{2}}~e^{-\mathbf{i} (1-\delta)\tau} 
+\int_{0}^{1} d\delta~ C^{-}_{n}~\delta^{n+\frac{1}{2}} ~e^{\mathbf{i} (1-\delta)\tau}
\Big] \nonumber \\
&=& -\mathbf{i}\epsilon^{2}~\frac{2t_B}{\tau^{\frac{3}{2}}}\Big[ e^{-\mathbf{i} \tau}\int_{0}^{\tau} dz~ \sum_{n=0}^{\infty}\frac{ C^{+}_{n}}{\tau^{n}}~z^{n+\frac{1}{2}}e^{\mathbf{i}~z}
+e^{i \tau} \int_{0}^{\tau} dz~ \sum_{n=0}^{\infty}\frac{ C^{-}_{n}}{\tau^{n}}~z^{n+\frac{1}{2}}e^{-\mathbf{i}~z}
\Big] \label{supp:sigK_int}
\end{eqnarray}
\end{widetext}
where $\tau=2t_B ~(t-t')$ ,$z=\delta~\tau$ and the subscript $l$ is dropped here for notational convenience. These integrations can be performed in the form of incomplete Gamma function $\Gamma (\alpha,z)$,
\begin{equation}
\int_{0}^{\tau} dz~z^{n+\frac{1}{2}}e^{\mathbf{i}~z} = - (-\mathbf{i})^{-n-\frac{3}{2}} \Big[     \Gamma(n+\frac{3}{2},-\mathbf{i} \tau)  -    \Gamma(n+\frac{3}{2},0)  \Big] 
 \label{incomplete_gamma}
\end{equation} 
which has its asymptotic form,
\begin{equation*}
\Gamma(n+\frac{3}{2},-\mathbf{i} \tau) = e^{\mathbf{i} \tau} (-\mathbf{i} \tau)^{n+\frac{1}{2}} \Bigg[ 1 +\frac{\mathbf{i}}{\tau} \Big(n+\frac{1}{2}\Big)  + \mathcal{O}\bigg(\frac{1}{\tau^{2}}\bigg) \Bigg] 
\end{equation*}
Using this and substituting the values of $C_{0}^{\pm}$, we obtain $\Sigma^{K}(t-t')$ which is given by equation (\ref{supp:sigk_full}). It is thus clear that the leading order power law tail in $\Sigma^{K}(t-t')$ goes as $|t-t'|^{-\frac{3}{2}}$ . 
\begin{widetext}
\begin{eqnarray}
\Sigma^K(t-t') &\sim &  -\mathbf{i} \epsilon^{2}~ t_B~\sqrt{\frac{\pi}{~|2t_B (t-t')|^{3}}} \Bigg( (-\mathbf{i})^{-\frac{3}{2}} e^{-\mathbf{i} 2t_B (t-t')} ~C^{+}_{0} + (\mathbf{i})^{-\frac{3}{2}}     e^{\mathbf{i} 2t_B (t-t')} ~ C^{-}_{0} \Bigg)+ \mathcal{O}\bigg(\frac{1}{\tau^{2}}\bigg) \nonumber \\
&\sim &-\mathbf{i} \epsilon^{2} ~\sqrt{\frac{2}{\pi~|2t_B (t-t')|^{3}}} \Bigg(  e^{-\mathbf{i} \left[2t_B (t-t') -\frac{3\pi}{4}\right]} \coth\left[\frac{2t_B-\mu}{2T}\right]   +    e^{\mathbf{i} \left[ 2t_B (t-t') -\frac{3\pi}{4}  \right] } \coth\left[\frac{-2t_B-\mu}{2T}   \right] \Bigg)+ \mathcal{O}\bigg(\frac{1}{\tau^{2}}\bigg)  \nonumber \\
\label{supp:sigk_full}
\end{eqnarray}
\end{widetext}
The leading order answer for $Im[\Sigma^{K}(t-t')]$ is then ,
\begin{widetext}
\begin{eqnarray}
\label{supp:sigK_leading}
Im[\Sigma^K(t-t') ]
&\sim &- \epsilon^{2} ~\sqrt{\frac{2}{\pi~|2t_B (t-t')|^{3}}} \Bigg(   \coth\left[\frac{2t_B-\mu}{2T}\right]   +     \coth\left[\frac{-2t_B-\mu}{2T}   \right] \Bigg) \cos   \left[ 2t_B (t-t') -\frac{3\pi}{4} \right] 
\end{eqnarray}
\end{widetext}
The expression , without the oscillation , is plotted in Fig. [1] as a solid line and matches with the numerically obtained plot very well.

Having shown the connection between the non-analyticity and power law tails, we now show that for the 2-site model , $G^{R}_{1,2}$ also has a $|t-t'|^{-\frac{3}{2}}$ tail . Similar considerations will apply for other $G^{R}_{i,j}$ and $G^{K}_{i,j}$ , both in the 2 site model and in the chain. Writing $\mathcal{D} = \mathcal{D}^{'} + \mathbf{i} \mathcal{D}^{''}$ the retarded Greens' function $G^{R}_{1,2}$ is given by,
\begin{eqnarray}
Re\left[ G^{R}_{1,2} \right] = -\frac{1}{g}~ \frac{\mathcal{D}^{'2}-\mathcal{D}^{''2}-1}{(\mathcal{D}^{'2}-\mathcal{D}^{''2}-1)^{2}+4\mathcal{D}^{'2}\mathcal{D}^{''2}} \nonumber \\
Im\left[ G^{R}_{1,2} \right] = -\frac{1}{g}~ \frac{-2\mathcal{D}^{'}\mathcal{D}^{''}}{(\mathcal{D}^{'2}-\mathcal{D}^{''2}-1)^{2}+4\mathcal{D}^{'2}\mathcal{D}^{''2}}
\end{eqnarray}
where
\begin{eqnarray}
\label{supp:Diag}
&&\mathcal{D}^{'} \nonumber  \\
&=& \frac{1}{g} \left[ w \bigg( 1- \frac{\epsilon^{2}}{2t_{B}^{2}} \bigg)+
\frac{\epsilon^{2}}{t_{B}}sign(\omega)\Theta\big(|\omega|-2t_{B}\big) \sqrt{\frac{\omega^{2}}{4t_{B}^{2}}-1}  \right] \nonumber \\
&&\mathcal{D}^{''} =  \frac{1}{g}  \frac{\epsilon^{2}}{t_{B}} \Theta\big(4t_{B}^{2}-\omega^{2}\big) \sqrt{1- \frac{\omega^{2}}{4t_{B}^{2}}}      
\end{eqnarray}
We note that $\mathcal{D}^{''} \sim \delta^{\frac{1}{2}}$ for $\omega = \pm 2t_B (1-\delta)$ while $\mathcal{D}^{'} \sim \left(1+C_{0}~ \delta^{\frac{1}{2}} \right)$ for $\omega = \pm 2t_B (1+\delta)$ and one needs to consider all these non-analytic pieces to calculate the power law tail. For Fourier transform of $Im[G^{R}_{1,2}]$ , the integration goes from $-2t_B < \omega < 2t_B$ and since $\mathcal{D}^{'}$ is smooth in this region, only non-analyticity of $\mathcal{D}^{''}$ leads to power law. One can then follow an argument similar to that of $\Sigma^{K}(t-t')$ and show that leading $\delta^{\frac{1}{2}}$ non-analyticity implies a power law $\sim |t-t'|^{-\frac{3}{2}}$.
For $Re[G^{R}_{1,2}]$, the argument is more complicated. The integral is broken into three ranges, (a) from $-\infty$ to $-2t_B$ , (b) from $2t_B$ to $\infty$ and (c) from $-2t_B$ to $2t_B$. Note that in the regions (a) and (b) the integrand can be expanded near $\omega = \pm 2t_B (1+\delta)$  as $\sum_{n=0}^{\infty} C^{\pm}_{n}~\delta^{\frac{n}{2}}$ coming from $\mathcal{D}^{'2}$ while the same in the region (c) goes as $\sum_{n=0}^{\infty} C^{\pm}_{n}~\delta^{n}$ near $\omega = \pm 2t_B (1-\delta)$. Then all the three integral in the ranges (a) , (b) and (c) have $\mathcal{O} \left(\tau^{-1} \right)$ coefficients which add up to zero and only integrals (a) and (b) contribute to obtain a finite $\mathcal{O} \left(\tau^{-\frac{3}{2}} \right)$ coefficient. Thus Fourier transform of $Re[G^{R}_{1,2}]$ also scales as $ |t-t'|^{-\frac{3}{2}}$ for $|t-t'| \rightarrow \infty$. We note that similar arguments can also be applied in analysis of $G^{K}$ to show its $ |t-t'|^{-\frac{3}{2}}$ decay in the long time limit.

\begin{figure*}[t]
\centering
\includegraphics[width=0.45\textwidth]{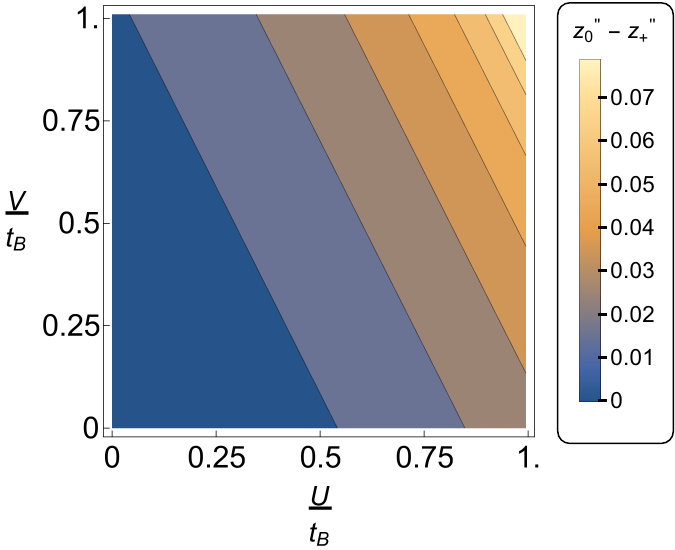} 
\includegraphics[width=0.45\textwidth]{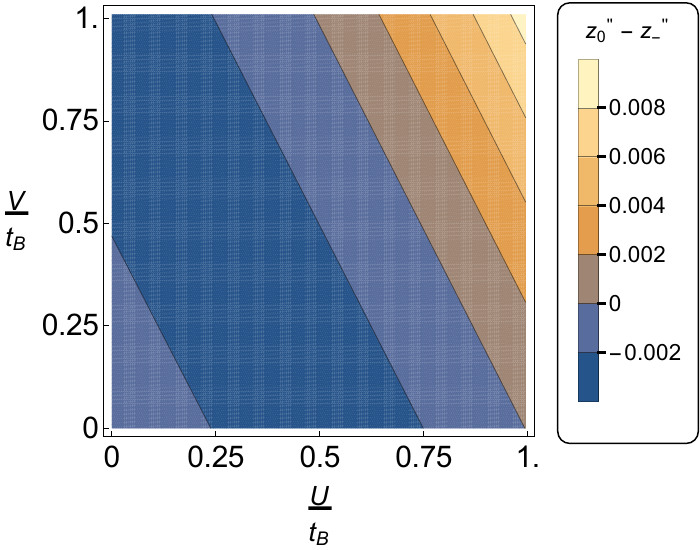}
\caption{The difference between the imaginary part of the pole for the non-interacting case  ($z_{0}^{''}$) and those of the interacting case ($z_{\pm}^{''}$) is plotted as a function of the interaction strength $U/t_B$ and $V/t_B$ for the 2-site model connected to two independent baths of same temperature $T=0.625t_B$ but different chemical potential $\mu_1=-2.5t_B$ and $\mu_2=-5.0t_B$ where we use $g=0.5t_B$ , $\epsilon=0.2t_B$. It shows that in the entire parameter space of the interaction strength , one of the two poles ($z_{+}$) of the interacting model always has smaller imaginary part than the non-interacting one. Hence the cross-over timescale from the exponential decay to power law tail in Greens' function and consequently in unequal time obsevables always shifts to larger $t-t'$ values as depicted in Fig. 6(c).     }
\label{figure:pole}
\end{figure*}
\section{ Exponential decay of current in a linear chain }
In section (IV) of the main text, we have derived the retarded and the Keldysh Greens' function for a linear chain, where each site is coupled to an independent bath with its own temperature and chemical potential. In section (V) , we found that when the baths have a common temperature but chemical potential of the bath varies linearly with lattice site no., the current in the links of the system show an exponential decay in space, away from the edges of the chain. In this appendix we use the analytic forms of the Greens' function in section (IV) and consider their low temperature form to explain the exponential decay of current. We will also show that the decay length is proportional to the temperature of the baths. 
To illustrate this, we use eqns. (14) and (16) of the main text to write,

\begin{widetext}
\begin{equation}
G^{K}_{l,l+1} =  \mathbf{i} \frac{\epsilon^{2}~J(w)}{g^{2}~|M_{N}|^{2}}  \Bigg(  M_{N-l}~M_{N-l-1}^{*}~ \sum_{j=1}^{l}|M_{j-1}|^{2} \coth \left[ \frac{w-\mu_{j}}{2T}\right] 
+ M_{l-1}~M_{l}^{*}~ \sum_{j=l+1}^{N}|M_{N-j}|^{2} \coth \left[ \frac{w-\mu_{j}}{2T}\right]
\Bigg) 
\label{supp:gk_l_l+1}
\end{equation}
\end{widetext}
where $M_{l}$ is now written in a form different from eqn (15) of the main text to facilitate the derivation. We now write,
\begin{equation}
M_{l}  = \frac{y^{l+1} - \frac{1}{y^{l+1}}}{2  \sqrt{\frac{D^{2}}{4}-1}}~~~~with~~~~~ y=\frac{\mathcal{D}}{2} + \sqrt{\frac{\mathcal{D}^{2}}{4}-1}   
\end{equation}
where $\mathcal{D}$ is given by eqn (\ref{supp:Diag}) of Appendix A. One can easily check that this definition of $M_{l}$ is equivalent to that in equation (15) of the main text. 

We now consider the low temperature limit , where $(\omega -\mu_{j}) >>  T~ ,\forall~\omega$ and $ \forall~j$ . In this case, $
\coth \left[ \frac{w-\mu_{j}}{2T}\right] \sim ~ 1+2~e^{-\left(\omega -\mu_{j}  \right)/T} =1+2 e^{-\left(\omega -\mu_{1} +\nu  \right)/T} e^{\nu~j/T} $ where $\mu_{l} = \mu_{1} + \nu ~(l-1)$ with $\nu= \frac{d\mu}{dx}$ setting the slope of linear variation of $\mu$ . Now the first term in the expansion of $\coth$ must sum up to zero , since this has no information about the variation of $\mu$ across the chain and current must be zero if all $\mu_{j} = \mu$ . We then focus on the second term. Writing $y=re^{\mathbf{i}\theta}$, we then get for this term,
\begin{widetext}
\begin{eqnarray}
\sum_{j=1}^{l}|M_{j-1}|^{2} \coth \left[ \frac{w-\mu_{j}}{2T}\right] &=& 2~ e^{-\left(\omega -\mu_{1} +\nu \right)/T}~ \sum_{j=1}^{l} \left[ r^{2j}  + r^{-2j} - 2~\cos(2j~\theta)   \right]~ e^{\nu~j/T} \nonumber \\
\sum_{j=1}^{l}|M_{N-j}|^{2} \coth \left[ \frac{w-\mu_{j}}{2T}\right] &=& 2~ e^{-\left(\omega -\mu_{1} +\nu  \right)/T}~ \sum_{j=1}^{l} \left[ r^{2(N-j+1)}  + r^{-2(N-j+1)} - 2~\cos\big(2(N-j+1)~\theta \big)   \right]~ e^{\nu~j/T} \nonumber
\end{eqnarray}
\end{widetext}
Now, we can have either $r>1$ or $r<1$ . We will first focus on the case $r>1$ and later show that similar argument works for $r<1$. Here, we are interested in the region of the chain far away from its boundary so that $r^{N}>> r^{l}$. 
The first series can then be summed (GP) to obtain terms like, $\left[\left( r^{2}e^{\nu/T} \right)^{l} -1\right]r^{2}e^{\nu/T}/  \left( r^{2}e^{\nu/T}-1 \right) + r \leftrightarrow e^{\pm \mathbf{i}\theta} $. One can then multiply with appropriate factors using equation (\ref{supp:gk_l_l+1}) to obtain $G^{K}_{l,l+1}$ whose leading order terms fall off exponentially $ \sim e^{\nu~l/T}$ where the other terms are suppressed by factor of $r^{2N}$ coming from $|M_{N}|^{2}$. This shows that $\xi = T~/~\nu$ and explains the linear scaling of $\xi$ with $T$ seen in Fig 3(f). One can similarly work out the other terms and the conclusion obtained above remains robust. If $r<1$, one can neglect $r^{i}$ in favour of $r^{-i}$ , but the final conclusion $I\sim e^{\nu~l/T}$ works out. Note that $\nu <0$ in our case, and hence this indicates a decay of current in space. 
A similar argument can be made for the density profile, except the term with 1 in the expansion of $\coth$ does not sum to zero. So $n(x) \sim n_{0} + n_{1} e^{\nu~l/T} $, i.e the variation around the constant value decays exponentially. 

\section{Effect of interaction on non-Markovian dynamics}
In section (VII) , we had shown that, within mean field theory , the exponential decay in the temporal profile of the Greens' functions of the interacting system is slower than that of the non-interacting system, and hence the time scale for cross-over from the "quasi-Markovian" to non-Markovian dynamics increases. We argued this based on the fact that imaginary part of the one of the poles of the Greens' function, which controls the decay rate, decreases with interaction. In Fig [\ref{figure:pole}] , we plot the differences in imaginary part of the pole between non-interacting and interacting systems as a function of $U$ and $V$ . We have considered a 2-site system with $g=0.5t_B$ , $\epsilon=0.2t_B$ , coupled to two baths with common temperature $T=0.625t_B$ , $\mu_1=-2.5t_B$ and $\mu_2=-5.0t_B$. We find that the difference $>0$ for all $U$ and $V$ for one of the poles , i.e, the non-interacting Greens' function decay faster. For the other pole , the difference is negative for small vales of $U$ and $V$ and becomes positive for larger values of $U$ and $V$ . We note that the long time decay (before power law tail) is governed by the smallest decay rate , and hence the cross-over time scale increases for all $U$ and $V$ .

\bibliographystyle{apsrev4-1}
\bibliography{Bathdynamics.bib}

\begin{thebibliography}{88}
\bibitem [{g0k()}]{g0k}%
  \BibitemOpen
  \href@noop {} {\ }\bibinfo {note} {Here, $G^{-1K}_{0}$ is a purely regulatory
  term governing intial condition of the system and can be neglected for steady
  state dynamics when the system is coupled to bath.}\BibitemShut {Stop}%
\bibitem [{pol()}]{pole0}%
  \BibitemOpen
  \href@noop {} {\ }\bibinfo {note} {$m=0$ corresponds to $\lambda = 0$ and in
  this case $\sinh \left[ (N+1) \lambda \right]/ \sinh \lambda \rightarrow
  (N+1)$. Hence $m=0$ does not correspond to a pole of the Greens'
  function.}\BibitemShut {Stop}%
\end{thebibliography}
\end{document}